\documentclass[aip,reprint]{revtex4-1}

\draft 

\usepackage{graphicx}
\usepackage{dcolumn}
\usepackage{bm}

\usepackage[utf8]{inputenc}
\usepackage[T1]{fontenc}
\usepackage{mathptmx}
\usepackage{etoolbox}
\usepackage{amsmath,amssymb,amsfonts,upref,endnotes}
\usepackage{amsthm}
\usepackage{xcolor}
\usepackage[caption=false]{subfig}

\usepackage[compact]{titlesec}        
\newtheorem*{remark}{Remark}

\begin{document}


\title{Tipping mechanisms in a carbon cycle model}

\author{Katherine Slyman}
\email{katherine.slyman@bc.edu}
\affiliation{ 
Department of Mathematics, Boston College, Chestnut Hill, MA 02467
}%
\author{Emmanuel Fleurantin}%
\author{Christopher K.R.T. Jones}
 \altaffiliation[Also at ]{RENCI, University of North Carolina at Chapel Hill, Chapel Hill, NC 27517}
\affiliation{
Department of Mathematics, George Mason University, Fairfax, VA 22030
}%

\date{\today}



\begin{abstract}
Rate-induced tipping (R-tipping) occurs when a ramp parameter changes rapidly enough to cause the system to tip between co-existing, attracting states, while noise-induced tipping (N-tipping) occurs when there are random transitions between two attractors of the underlying deterministic system. This work investigates R-tipping and N-tipping events in a carbonate system in the upper ocean, in which the key objective is understanding how the system undergoes tipping away from a stable fixed point in a bistable regime. While R-tipping away from the fixed point is straightforward, N-tipping poses challenges due to a periodic orbit forming the basin boundary for the attracting fixed point of the underlying deterministic system. Furthermore, in the case of N-tipping, we are interested in the case where noise is away from the small noise limit, as it is more appropriate for the application. We compute the most probable escape path (MPEP) for our system, resulting in a firm grasp on the least action path in an asymmetric system of higher scale. Our analysis shows that the carbon cycle model is susceptible to both tipping mechanisms when using the standard formulations.
\end{abstract}


\maketitle 

\begin{quotation}
Tipping dynamics can best be understood in bistable systems, where qualitatively different stable states coexist for a given set of parameters within the system \cite{Kaszas.2019}. Many systems exhibit these multiple, alternative stable states where underlying dynamics, parameter values, or stochasticity can cause the abrupt shift from one stable state to another, such that the system cannot continuously track the stable state. 
We focus on a marine carbonate cycle model proposed by \citet{Rothman.2019}. This is a complex system not uniquely affected by only one tipping mechanism, and in fact, exhibits bifurcation, rate, and noise-induced tipping. The original work of Rothman looks at the system in an excitable regime, while this work considers a standard formulation of rate-induced tipping as well as noise-induced tipping. The guiding principles of this work come from geometric dynamical systems methods along with numerical simulations for corroboration and visualization. This approach implies we juggle between two points of view: the dynamical systems and stochastic perspectives. 

Our focus is on the escape from a fixed point through an unstable periodic orbit, both due to the ramping of an external carbon dioxide injection rate and additive white noise on the system's state variables. We find that the dynamical system is forward threshold unstable, which implies the system is susceptible to rate-induced tipping. Additionally, within the stochastic version of the system, for noise levels away from the small noise limit, Monte Carlo simulations show there is a specific region on the unstable periodic orbit where trajectories escape, rather than the cycling behavior predicted for vanishingly small noise \citep{day_exit_1996}. We find a subset of the unstable manifold, in the Freidlin-Wentzell system, of the stable fixed point from the deterministic system that is comprised of local minimizers, and plays a key role in determining the window of escape on the periodic orbit. Using the Maslov Index and the Onsager-Machlup functional, we are able to bracket the window of escaping trajectories.

\end{quotation}

\section{Introduction}
The ocean is a crucial component in Earth's climate system. Through an interplay of different processes, it redistributes heat and carbon across the planet: the carbon cycle describes these processes that exchange carbon between the ocean, atmosphere, and the continents \cite{williams2011_book1, archer2011warming_book2}. 
The ocean absorbs much of the carbon dioxide (CO$_2$); it is the largest sink of CO$_2$ on the planet, absorbing up to a third of anthropogenic CO$_2$ emissions, and actually, contains on the order of fifty times the amount of carbon than that in the atmosphere\cite{williams2011_book1}.
Carbon dioxide in the atmosphere reacts with seawater to form bicarbonate and carbonate ions. The sum of these ions, together with dissolved CO$_2$, is known as dissolved inorganic carbon, but more than 99\% of dissolved inorganic carbon are the bicarbonate (HCO$^-_3)$ and carbonate (CO$^{2-}_3)$ ions \cite{williams2011_book1}. 

The amount of CO$_2$ in the atmosphere is on the rise, and rising amounts of atmospheric CO$_2$ lead to ocean acidification \cite{archer2011warming_book2, NationalOceanicandAtmosphericAdministration.2013}. This process reduces the ocean's ability to absorb carbon, resulting in more carbon staying in the atmosphere. Additionally it decreases the pH levels of the ocean, making it difficult for marine species and ecosystems to survive \cite{archer2011warming_book2}. These species will either have to adapt to the changing ocean or risk extinction. Understanding the carbon cycle is of vital importance for our society as the preservation of these ecosystems and organisms are tied to fishing, industrial materials, and tourism industries \cite{NationalOceanicandAtmosphericAdministration.2013}.

Rothman \cite{Rothman.2019} presents a dynamical system encompassing the important features of the marine carbonate cycle. The system is composed of two ordinary differential equations, which track the concentrations of dissolved inorganic carbon and carbonate ions. The main parameter of interest is the rate of injection of external CO$_2$ (where external sources are classified as human activities and volcanic emissions), and then studying the model as this rate increases. He determined that when the external CO$_2$ injection rate increased and crossed a critical threshold value where the periodic orbits collided in a saddle-node bifurcation, both the concentrations of the dissolved inorganic carbon and the carbonate ions, when near a stable state, can be excited into a vestigial oscillatory regime, and in time, relax back to the stable state. This oscillatory behavior corresponds to significant disruptions of the ancient earth system, including mass extinction events. 

After analyzing the model, Rothman examines the implications of these findings for interpreting past disruptions in the geochemical record, as well as forecasts the carbon cycle's response to modern human-driven disturbances. In a recent MIT news article \cite{Chu.2019}, Rothman hypothesized that based on the rapid rate CO$_2$ is entering the atmosphere, the planet is likely to reach a critical threshold by the end of the century, with bad consequences and the potential culmination in Earth's sixth mass extinction event.

The results of Rothman's work, along with the physical importance of the ocean, demonstrate the need for understanding the carbon cycle and its underlying mechanisms. We use \cite{Rothman.2019} as our motivation and study the system's susceptibility to tipping, where we broadly define tipping as the rapid, and often irreversible, change in the state of a system \cite{ashwin_parameter_2017}. There are three main mechanisms for tipping in dynamical systems: bifurcation-induced, rate-induced, and noise-induced \cite{ashwin_tipping_2012}, and they have been classified according to whether they involve a bifurcation in the system (B-tipping), a parameter shift (R-tipping), or the addition of random fluctuations (N-tipping). Unlike B-tipping, with R-tipping there is no bifurcation in the system to explain the change of stability of a fixed point. B-tipping and R-tipping both occur in deterministic systems, while N-tipping requires a stochastic component.

The original work of Rothman \cite{Rothman.2019} can be viewed as the demonstration of a side effect of a bifurcation. Since the parameter is fixed, or only changing slowly, it is not a case of R-tipping, even though it may seem that way as the parameter is a rate itself. Rothman's key message relates to the system being in a regime that is excitatory and not bistable. 

For the work in this paper, we focus completely on the bistable regime and the types of tipping that can occur from the stable state to a large oscillation. It is reasonable to consider the external injection of CO$_2$ ramping up in time as, for instance, human emissions continue to rapidly increase. A key question becomes: for external inputs of the CO$_2$ less than the critical threshold value, and hence squarely in the bistable regime, can the system exhibit tipping if the change of the parameter value is fast enough in time?

The work of \citet{Rothman.2019} was a deterministic study. It is reasonable to also consider random fluctuations acting to the carbon cycle system. Most probable path theory has its origins in such settings where stochastic perturbations are added to a deterministic setting. 
Random events such as natural disturbances (e.g. hurricanes, droughts) and ecosystem disturbances (e.g. insect outbreaks) can affect carbon uptake and release. In this particular context, noise may be small but not vanishingly so. This motivates the second focus of this work: finding which parts of the theory of large deviations hold true for small, but not vanishingly small, levels of noise, and how to best capture the behavior exhibited by the system for these noise strengths. Related work was carried out by \citet{duan1,duan2}, who study the same underlying model, but with multiplicative noise.

This paper is organized as follows. Section \ref{Sec: OG Model} begins with a brief model description and looking at the regime of study. The regime chosen is bistable and exhibits a stable fixed point and a stable periodic orbit, separated by an unstable periodic orbit. Section \ref{sec:Rtip} considers the deterministic system and the possibility of rate-induced tipping away from the stable fixed point when the system's parameter representing the injection of external CO$_2$ is time-dependent. After considering basins of attraction of the stable fixed point and the stable periodic orbit, and choosing a sufficient parameter shift, the system is susceptible to R-tipping away from the stable fixed point due to its forward threshold instability. 

Section \ref{sec:noise} considers the effects of stochasticity on the state variables of the model in which the focus is N-tipping from the stable fixed point to the stable periodic orbit. While demonstrating the system is susceptible to R-tipping is straightforward, N-tipping poses some challenges. The emphasis of this work is finding local minimizers of the Freidlin-Wentzell functional on the semi-infinite time domain $t \in (-\infty,t_f]$, where trajectories reach the unstable periodic orbit of the underlying deterministic system at $t=t_f$. Since ultimately we are looking for global minimizers for the small, but not vanishingly small noise regime, the extra step needed is the use of the Onsager-Machlup functional \cite{OM_Durr_Bach}, as a perturbation of the Freidlin-Wentzell functional, on the set of minimizers found. This analysis of the stochastic system is largely based on the analysis and the methods used in \citet{Emman_IVDP}; since the goal is to tip from a stable fixed point to an unstable periodic orbit, as the system would follow the deterministic flow once it reaches the basin boundary. Section \ref{sec:discussion} finishes with final conclusions and discussion.

We stress that this work is to study rate and noise-induced tipping within this system in their most standard formulation. When looking at R-tipping, we take the canonical rate function and similarly when looking at N-tipping, we take equal, additive noise on the state variables. We note that this analysis was physically motivated but we focus on the mathematical approaches needed to analyze the complex behavior.

\section{Modeling the marine carbonate cycle} \label{Sec: OG Model}

\subsection{Model setup}

Rothman \cite{Rothman.2019} creates a simple dynamical system model of the marine carbonate cycle given by
\begin{equation}
\begin{aligned}
 \dot{c} &= F(c,w)=f(c)[\mu(1-bs(c,c_p)-\theta \bar{s}(c,c_x)-\nu)+w-w_0], \\
 \dot{w} &= G(c,w)=\mu[1-bs(c,c_p)+\theta \bar{s}(c,c_x)+\nu]-w+w_0,
\end{aligned}
\label{EQ: CC short}
\end{equation}
where $c$ is the carbonate ion concentration and $w$ is the dissolved inorganic carbon concentration. The functions $f,s,\bar{s}$ model the buffering, burial, and respiration of the concentration of carbonate ions, and are explicitly given by the sigmoidal functions
\begin{equation}
\begin{aligned}
f(c)&=\frac{f_0 c^{\beta}}{c^{\beta}+c_f^{\beta}},\\
s(c,x)&=\frac{c^{\gamma}}{c^{\gamma}+x^{\gamma}},\\
\bar{s}(c,x)&=1-s(c,x).
\end{aligned}
\label{EQ: CC funcs}
\end{equation}

\noindent This model assumes a well-mixed ocean, implying the timescales considered must be larger than approximately 1000 years, which are the timescales necessary for global ocean mixing: in \eqref{EQ: CC short}, time is nondimensionalized by $t=t/\tau_w$, where $\tau_w=1000$. The parameters that differ from that of the supplementary material of \citet{Rothman.2019}, along with their descriptions, values we assume, and units, are listed in Table \ref{Table:CCvalues}. For a more detailed analysis or the derivation and explanation of the model and parameters, refer to \citet{Rothman.2019}.

In \citet{Rothman.2019}, the parameters $\mu, b, \theta, c_x, c_p$, and $\nu$ were of interest, and the remaining five parameters were set to correspond to the chemical equilibrium or the properties of the modern ocean, to maintain realism. For the purposes of this work, the parameters of focus are $c_x$ and $\nu$. The parameter $c_x$ is chosen to ensure a bistable regime, as shown in Figure \ref{fig:CC_2}.  Then, in Section \ref{sec:Rtip}, instead of considering fixed $\nu$, we consider $\nu$ as a time-dependent parameter, as it represents an external CO$_2$ injection rate.

\begin{table}[!h]
\caption{Parameter values of the marine carbonate cycle system \eqref{EQ: CC short} and their descriptions. We include those which differ from that of the supplementary material found in \citet{Rothman.2019}. All other parameter values not listed can be found in the aforementioned work.}
\label{Table:CCvalues}
\begin{tabular}{llll}
\hline
Parameter    & Value  & Units & Description \\ \hline
$c_x$ & 58 & $\mu$ \text{mol kg}$^{-1}$ & \text{crossover} [CO$_3^{2-}$] \text{ (respiration)}    \\
$\nu$ & 0.1 & - & \text{CO}$_2$ \text{injection rate}  \\
\hline
\end{tabular}
\vspace*{-4pt}
\end{table}

\subsection{Orbits of interest}
The fixed points of \eqref{EQ: CC short} occur where $\dot{c}=\dot{w}=0$, implying the unique steady-state solution $(c^*,w^*)$ is
\begin{equation}
\begin{aligned}
 c^*&=c_p(b-1)^{-1/\gamma},\\
 w^*&=w_0 + \mu \left(\theta+\nu-\frac{\theta c_p^{\gamma}}{c_p^{\gamma}+(b-1)c_x^{\gamma}}\right).
\end{aligned}
\label{EQ: CC FP}
\end{equation}

\noindent Linearizing \eqref{EQ: CC short} about the fixed point $(c^*,w^*)$ from \eqref{EQ: CC FP} results in the Jacobian matrix
\begin{equation}
J(c^*,w^*)=\begin{pmatrix}
F_c(c^*,w^*) & F_w(c^*,w^*)  \\
G_c(c^*,w^*)  & G_w(c^*,w^*) 
\end{pmatrix}.
\label{EQ: Jacobian 2d}
\end{equation}

\noindent The eigenvalues of $J(c^*,w^*)$ are complex conjugate pairs, and as $c_x$ increases, these eigenvalues cross the imaginary axis, where this crossing of the imaginary axis implies the system undergoes a Hopf bifurcation.

A bifurcation analysis is needed to further understand the possible solution behaviors that \eqref{EQ: CC short} can exhibit; \citet{Rothman.2019} produces a bifurcation diagram. The parameter values we use match this analysis, except for $\nu$, where $\nu=0$ in the original work. We however set $\nu=0.1$. We find using MatCont \cite{dhooge2003matcont}, a Matlab software project that provides a toolbox for the numerical continuation and bifurcation study of continuous and discrete parameterized dynamical systems, the bifurcation diagram for this parameter regime is nearly identical, and thus we omit it from this work. 

\begin{figure*}
  \centering
  \subfloat[]{\label{fig:CC_1}\includegraphics[scale=.35]{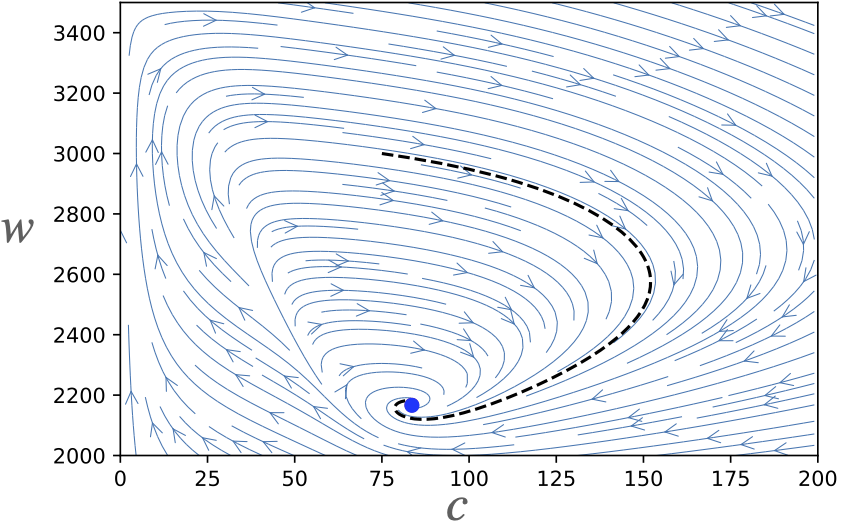}}
  \subfloat[]{\label{fig:CC_2}\includegraphics[scale=.35]{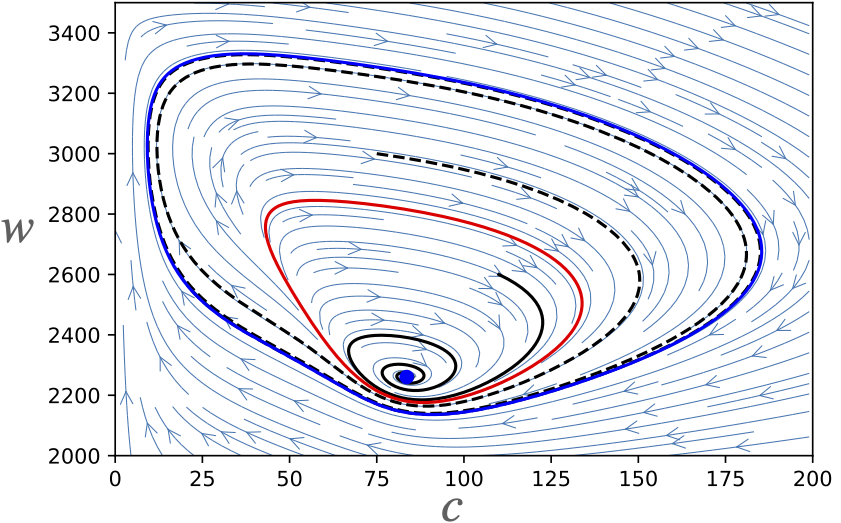}}
  \subfloat[]{\label{fig:CC_3}\includegraphics[scale=.35]{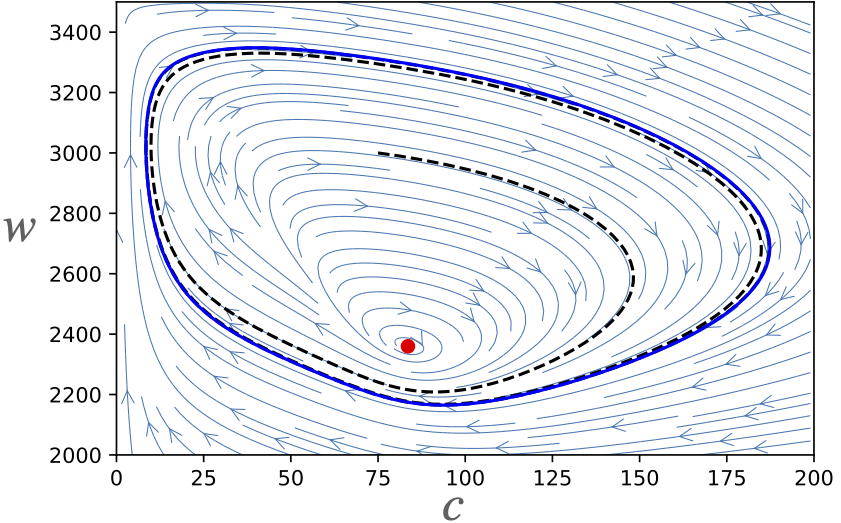}}
 \caption{Phase planes of system \eqref{EQ: CC short} with all parameters set as in Table \ref{Table:CCvalues} and the supplementary material of \citet{Rothman.2019}, with varying $c_x$. Stable states are in blue and unstable states are in red. Black solid and dashed curves represent single solution trajectories. (a) $c_x=55$: There is a stable fixed point. (b) $c_x=58$: There is a stable periodic orbit and a stable fixed point separated by an unstable periodic orbit. (c) $c_x=65$: There is a stable periodic orbit and an unstable fixed point.}
   \label{FIG:CCbehaviors}
\end{figure*}

There are three different system behaviors dependent on the parameter regime chosen, within realistic values of $c_x$. It follows that there can be (i) a stable fixed point, (ii) an unstable fixed point and stable periodic orbit, or (iii) a stable fixed point and a stable periodic orbit separated by an unstable periodic orbit. See Figure \ref{FIG:CCbehaviors} for an illustration of the phase planes of these three dynamical regimes. In Sections \ref{sec:Rtip} and \ref{sec:noise}, the focus is on the regime in which there is bistability.

This bistable regime, where the parameter values are set as in Table \ref{Table:CCvalues} and the supplementary material of \citet{Rothman.2019}, corresponds to Figure \ref{fig:CC_2}. Let $z^*$ denote the fixed point $(c^*,w^*)$, let $\Gamma_u$ denote the unstable periodic orbit, and let $\Gamma_s$ denote the stable periodic orbit. The precise values of the following computations are rounded for ease of reading.

The fixed point in the bistable regime is given by
\begin{equation}
z^*=(83.581, 2260.290), 
\label{EQ: CC zstar}
\end{equation}
and the linearzation at \eqref{EQ: CC zstar} is 
\begin{equation}
J(z^*)=\begin{pmatrix}
0.087 & 0.519 \\
-18.114 & -1
\end{pmatrix}.
\label{EQ: Jacobian 2d numbers}
\end{equation}
The eigenvalues of the Jacobian matrix \eqref{EQ: Jacobian 2d numbers} are given by
\begin{equation}
\lambda_{1,2}=-0.456437  \pm 3.02052i,
\end{equation}
implying the fixed point \eqref{EQ: CC zstar} is a stable spiral. In the two-dimensional system, $z^*$ has a two-dimensional stable subspace and $\Gamma_u$ has a one-dimensional unstable subspace. Floquet multipliers are used to find the stability of the periodic orbits.

\section{The switch to a time-varying parameter}
\label{sec:Rtip}
We introduce an external input, or parameter shift, $\nu_r(t)$ to replace the fixed parameter $\nu$, which allows the external CO$_2$ injection rate to vary in time. This substitution changes \eqref{EQ: CC short} to a nonautonomous system of the form
\begin{equation}
\begin{aligned}
\dot{c} &= f(c)[\mu(1-bs(c,c_p)-\theta \bar{s}(c,c_x)-\nu_r(t))+w-w_0], \\
\dot{w} &= \mu[1-bs(c,c_p)+\theta \bar{s}(c,c_x)+\nu_r(t)]-w+w_0.
\end{aligned}
\label{EQ: nonaut CC}
\end{equation}

To understand the notions of R-tipping within this system, we first introduce needed notation and definitions from \citet{wieczorek2023rate}. Define the autonomous frozen system as \eqref{EQ: nonaut CC} when $\nu_r(t)=\nu$ is fixed at a specific value. We assume $\nu_r(t)$ will be bi-asymptotically constant, and gradually transition from $\nu_-$ to $\nu_+$ in time, where we call $\nu_-$ the past limit state and $\nu_+$ the future limit state. The past (future) limit state is just the autonomous frozen system when $\nu_r(t)=\nu_- (\nu_+).$

Suppose we start at a stable state at $\nu_-$. If this base state lies in the basin of attraction of a different stable state at $\nu_+$, then the system is forward threshold unstable. We denote the basin of attraction of a stable state, $\alpha$, at the external input value, $\lambda$, by $\mathbb{B}(\alpha,\lambda)$. See Definition 4.6(a) of \citet{wieczorek2023rate} for the formal definition of forward threshold instability. From the work of \citet{Kiers.2019, wieczorek2023rate}, a system is susceptible to R-tipping if the system is ever forward threshold unstable. Thus we have sufficient criteria for determining if R-tipping occurs in a system.

\subsection{The system's susceptibility to R-tipping} \label{Sec: chp2 rate p1}

We create a monotonically increasing, bi-asymptotically constant external input $\nu_r(t)$ given by
\begin{equation}
\nu_r(t)=\nu_-+\frac{\nu_+-\nu_-}{2}(1+\tanh(rt)),
\label{EQ: CC ramp gen}
\end{equation}
in which $\nu_r(t)$ will gradually transition from $\nu_-$ to $\nu_+$. The parameter values were chosen such that there is a stable fixed point and a stable periodic orbit, separated by an unstable periodic orbit. The unstable periodic orbit forms the boundary between the basins of attraction for the stable fixed point and the stable limit cycle. Consider the autonomous frozen system at both the past and future limit states. Let $p_-, \Gamma_{u^-}$, and $\Gamma_{s^-}$ represent the stable fixed point, unstable periodic orbit, and the stable periodic orbit at the past limit state, $\nu_-$. Let $p_+, \Gamma_{u^+}$, and $\Gamma_{s^+}$ represent the stable fixed point, unstable periodic orbit, and the stable periodic orbit at the future limit state, $\nu_+$. 

The system given by \eqref{EQ: nonaut CC} will be susceptible to rate-induced tipping if it is forward threshold unstable:  $p_-$ lies in the basin of attraction of $\Gamma_{s^+}$.
We choose $\nu_-=0$, signifying the possibility of the external CO$_2$ injection rate having no contribution.

\begin{figure}[ht]
  \centering
  \subfloat[]{\includegraphics[width=0.24\textwidth]{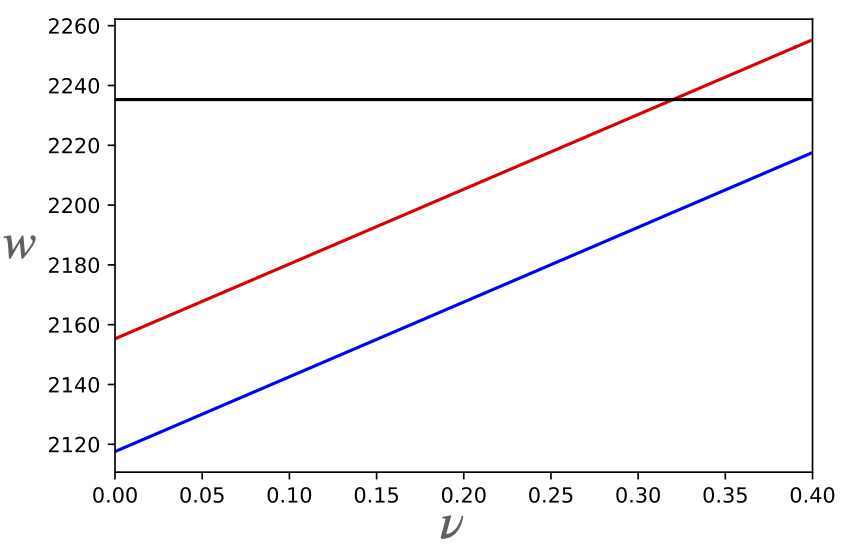}\label{fig:vchart}}
  \subfloat[]{\includegraphics[width=0.24\textwidth]{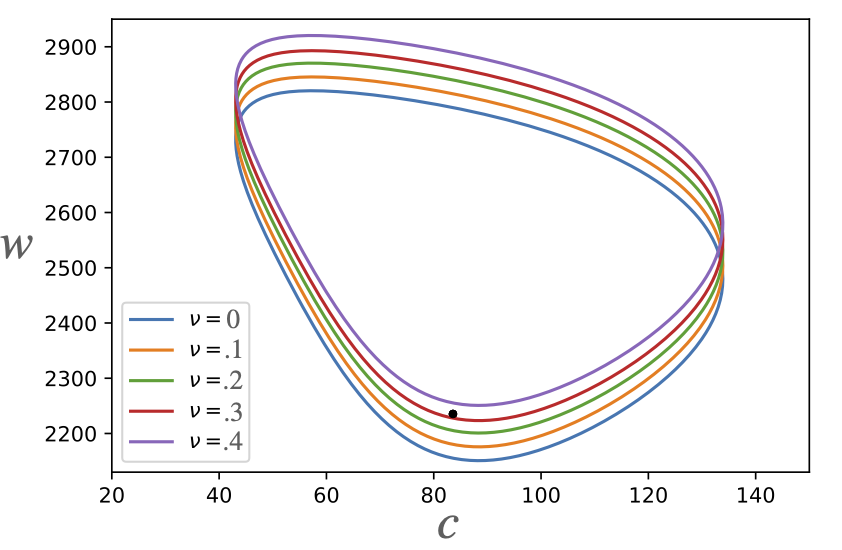}\label{fig:basins}}
  \caption{(a) The $w$ value of $v_-$ (black) overlaid with the $w$ value at $c=c^*$ for the unstable periodic orbit (red) and the stable periodic orbit (blue). (b) The stable fixed point $p_-$ (black circle) of \eqref{EQ: nonaut CC} at the past limit state $\nu_-=0$, plotted against the unstable periodic orbit $\Gamma_{u^+}$ at the future limit state $\nu_+$ for $\nu_+ \in [0,.1,.2,.3,.4]$ . Crossing of $\Gamma_u$ occurs for $(p_-,0) \in \mathbb{B}(\Gamma_{s^+},0.4)$, implying there will be rate-induced tipping away from $p_-$.}
   \label{FIG:rate basin}
\end{figure}

Recall the general formula of the stable fixed point given in \eqref{EQ: CC FP}. Notice that $c^*$ is independent of $\nu$, and therefore to verify the condition of forward threshold instability, we need only to check the value of $w^*$ of $p_-$ with the value of $w$ at $c=c^*$ on $\Gamma_u^+$. This calculation is performed numerically, as seen in Figure \ref{fig:vchart}. At $\nu_+ \approx 0.33$, we see $(p_-,\nu_-) \in \mathbb{B}(\Gamma_{s^+},\nu_+)$, implying the system is forward threshold unstable and the marine carbon cycle is susceptible to rate-induced tipping away from $p_-$. 

For further clarity and to better depict the notion of forward threshold instability, refer to Figure \ref{fig:basins}. In this figure, the past limit state is set at $\nu_-=0$ and we plot the corresponding stable fixed point $p_-$. The unstable periodic orbit $\Gamma_{u^+}$ is shown for four different possible values of the future limit state, $\nu_+$. For rate-induced tipping to occur, it is necessary that $(p_-,\nu_-) \in \mathbb{B}(\Gamma_{s^+},\nu_+)$, implying that $p_-$ must be on the outside of $\Gamma_{u^+}$. We see that for $\nu=0.4$, this transition occurs and the carbon cycle model will be susceptible to rate-induced tipping away from $p_-$. 

\subsection{Numerical results}
From the analysis in Section \ref{Sec: chp2 rate p1}, one option for the external input introduced in \eqref{EQ: CC ramp gen} is
\begin{equation}
\nu_r(t)=0.2(1+\tanh(rt)),
\label{EQ: CC ramp}
\end{equation}
which gradually transitions from $\nu_-=0$ to $\nu_+=0.4$ in time. To numerically simulate solutions of \eqref{EQ: nonaut CC}, we convert \eqref{EQ: nonaut CC} to an autonomous system. We augment the system by introducing a new variable, $y = t$, and making the corresponding substitutions and differentiating, results in the three-dimensional autonomous system given by

\begin{equation}
\begin{aligned}
\dot{c} &= f(c)[\mu(1-bs(c,c_p)-\theta \bar{s}(c,c_x)-\nu_r(y))+w-w_0], \\
\dot{w} &= \mu[1-bs(c,c_p)+\theta \bar{s}(c,c_x)+\nu_r(y)]-w+w_0, \\
\dot{y}&=1.
\end{aligned}
\label{EQ: aut CC}
\end{equation}
We note while this method is well-suited for numerical computations, it does not have the underlying structure to approach the problem from a dynamical systems perspective; there are no fixed points as $\dot{y}$ will never equal zero. To further study the rate-induced tipping problem in the dynamical systems framework, compactification \cite{wieczorek_compactification_2021} is preferred.

\begin{figure}[ht]
  \centering
  \subfloat[]{\includegraphics[width=0.23\textwidth]{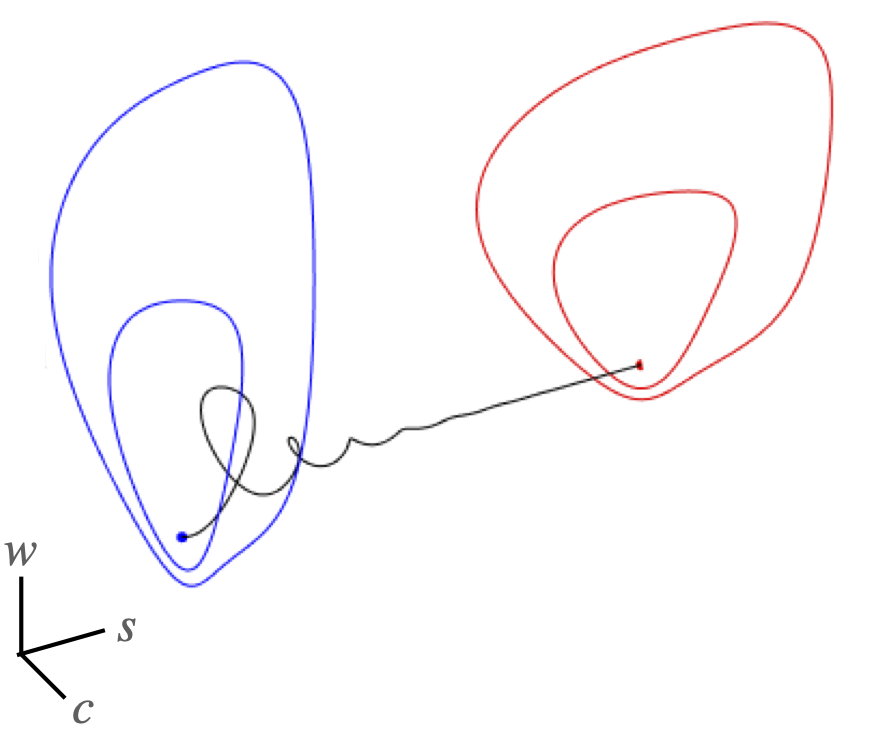}}
  \subfloat[]{\includegraphics[width=0.23\textwidth]{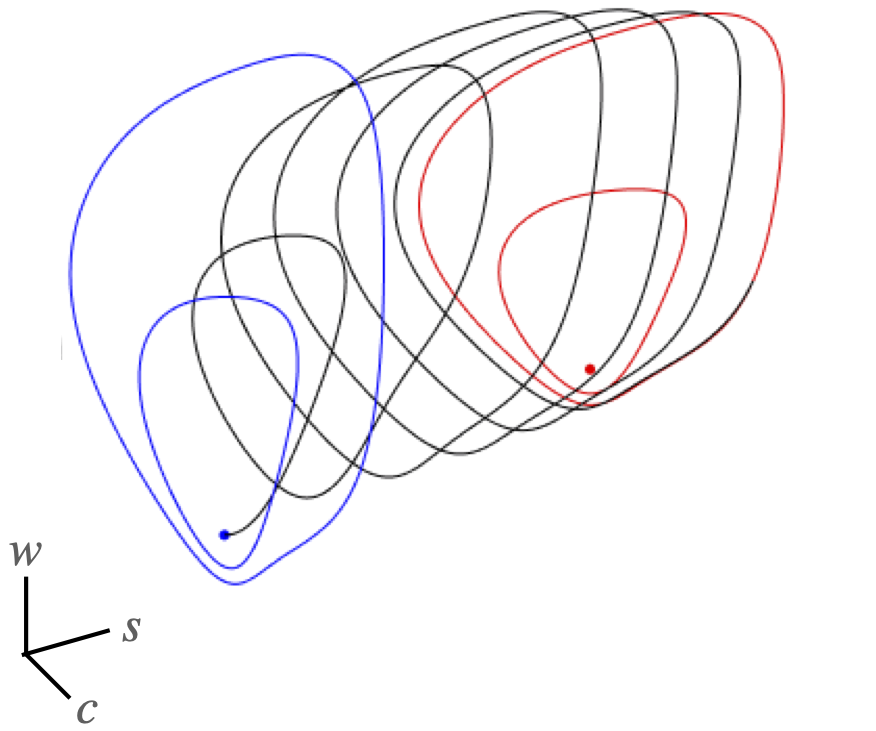}}
  \caption{Solution behaviors (black) of system \eqref{EQ: aut CC} for varying values of $r$, where the past limit system is in blue and the future limit system is in red. The system is initialized at $p_-$ (blue circle). (a) $r=1.5$: no tipping occurs as the solution end-point tracks $p_+$. (b) $r=1.62$: Rate-induced tipping occurs away from $p_-$ as the solution escapes the basin of attraction of $p_-$ and relaxes to $\Gamma_{s^+}$.}
   \label{FIG:CCrtip3d}
\end{figure}

\begin{figure}[ht]
  \centering
  \subfloat[]{\includegraphics[width=0.23\textwidth]{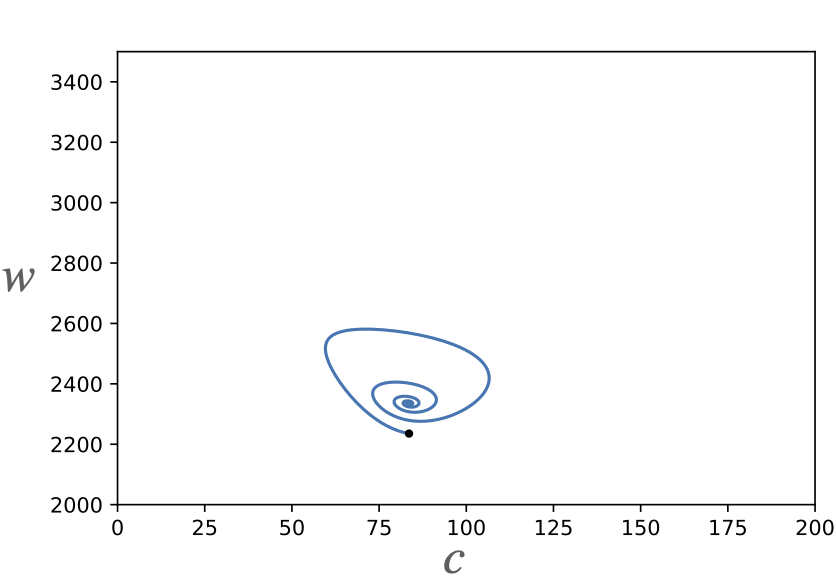}}
  \subfloat[]{\includegraphics[width=0.23\textwidth]{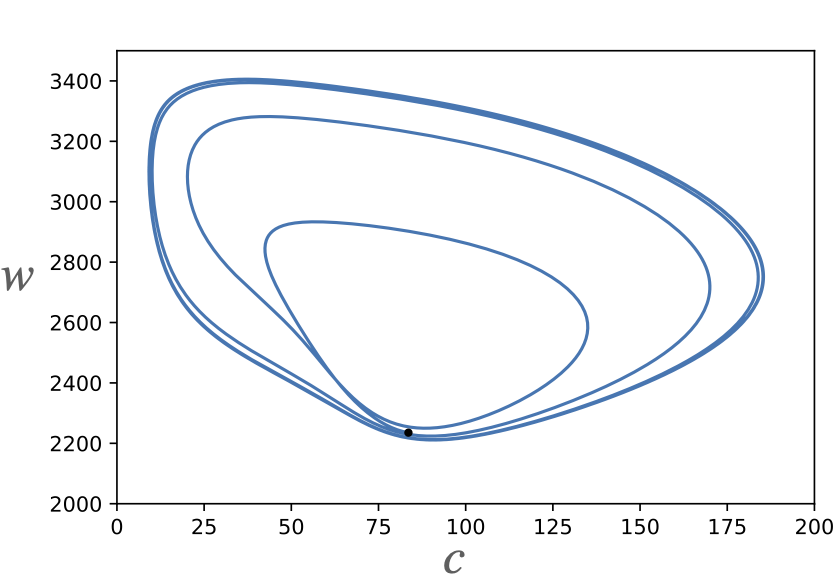}}
  \caption{Solution behaviors (blue) of system \eqref{EQ: aut CC} for varying values of $r$. The system is initialized at $p_-$ (black circle). (a) $r=1.5$: No tipping occurs as the solution end-point tracks $p_+$. (c) $r=1.62$: Rate-induced tipping occurs away from $p_-$ as the solution escapes the basin of attraction of $p_-$ and relaxes to $\Gamma_{s^+}$.}
   \label{FIG: CC rtip pics}
\end{figure}

Initializing \eqref{EQ: aut CC} at $p_-$, and numerically simulating \eqref{EQ: aut CC} for increasing values of $r$ demonstrates how solution behaviors change as $r$ increases. In Figure \ref{FIG:CCrtip3d}, notice that for sufficiently large values of $r$ the system tips, while for sufficiently small values of $r$ the system could continually readjust to the changing fixed point. This shows that the rate at which $\nu$ is varied determines whether the system tips, implying the existence of a critical rate, $r=r_c$, and above which tipping occurs. See Figure \ref{FIG: CC rtip pics} for a two-dimensional projection of the solution behaviors.

The main goal of this section was to show that the system is susceptible to rate-induced tipping, not to find the exact $r_c$, though at $r=r_c$, there is an orbit connecting the stable fixed point $p_-$ to the unstable periodic orbit $\Gamma_{u^+}$. 

In short, this section demonstrates that if the CO$_2$ injection rate $\nu$ is a time-dependent parameter and it increases fast enough, the system crosses over from the basin of attraction of the fixed point to that of the stable periodic orbit. Therefore, the system can tip between states before the actual loss of stability of the fixed point. Physically, this implies that the amounts of carbonate and dissolved inorganic carbon could change drastically in a relatively short time, affecting the oceans and their current balance. 

\section{The stochastic version of the model} \label{sec:noise}

We have studied the susceptibility of system \eqref{EQ: CC short} to R-tipping, and to continue the investigation, we now move to study its susceptibility to N-tipping by considering additive white noise on the dynamics of both $c$ and $w$. The approach used to study this phenomenon is largely based on the tools developed by \citet{Emman_IVDP}. While we summarize and provide the main details of the approach, we encourage the reader to see this work for the finer details of the analysis. 

We remain in the bistable regime, with the parameters set as in Table \ref{Table:CCvalues} and the supplementary material of \citet{Rothman.2019}, where $z^*, \Gamma_u, \Gamma_s$ are the stable fixed point, unstable periodic orbit, and stable periodic orbit respectively.

Consider the stochastic differential equation of the form
\begin{equation}
    dz=P(z) dt + \epsilon \Sigma dW.
\end{equation}
This SDE has underlying deterministic dynamics given by $\dot{z}=P(z)$, noise strength $\epsilon$, and noise structure $\Sigma$, which is an $n \times n$ matrix. For mathematical simplification and to really study the affect of additive noise, we let $\Sigma=I$, where $I$ is the $n \times n$ identity matrix. The chosen structure assumes equal noise strengths on the state variables, but different Weiner processes.

For the carbon cycle, $z=(c,w)$ and $P= (F,G)$ where $F,G$ are as given in \eqref{EQ: CC short}. For additional clarity, we note the stochastic version of \eqref{EQ: CC short} is given by
\begin{equation}
\begin{aligned}
 dc&=F(c,w)dt+\epsilon dW_1,\\
dw&=G(c,w)dt+\epsilon dW_2.
\end{aligned}
\label{EQ:CC stoch}
\end{equation}

\begin{figure}[ht]
\centering
\includegraphics[scale=.18]{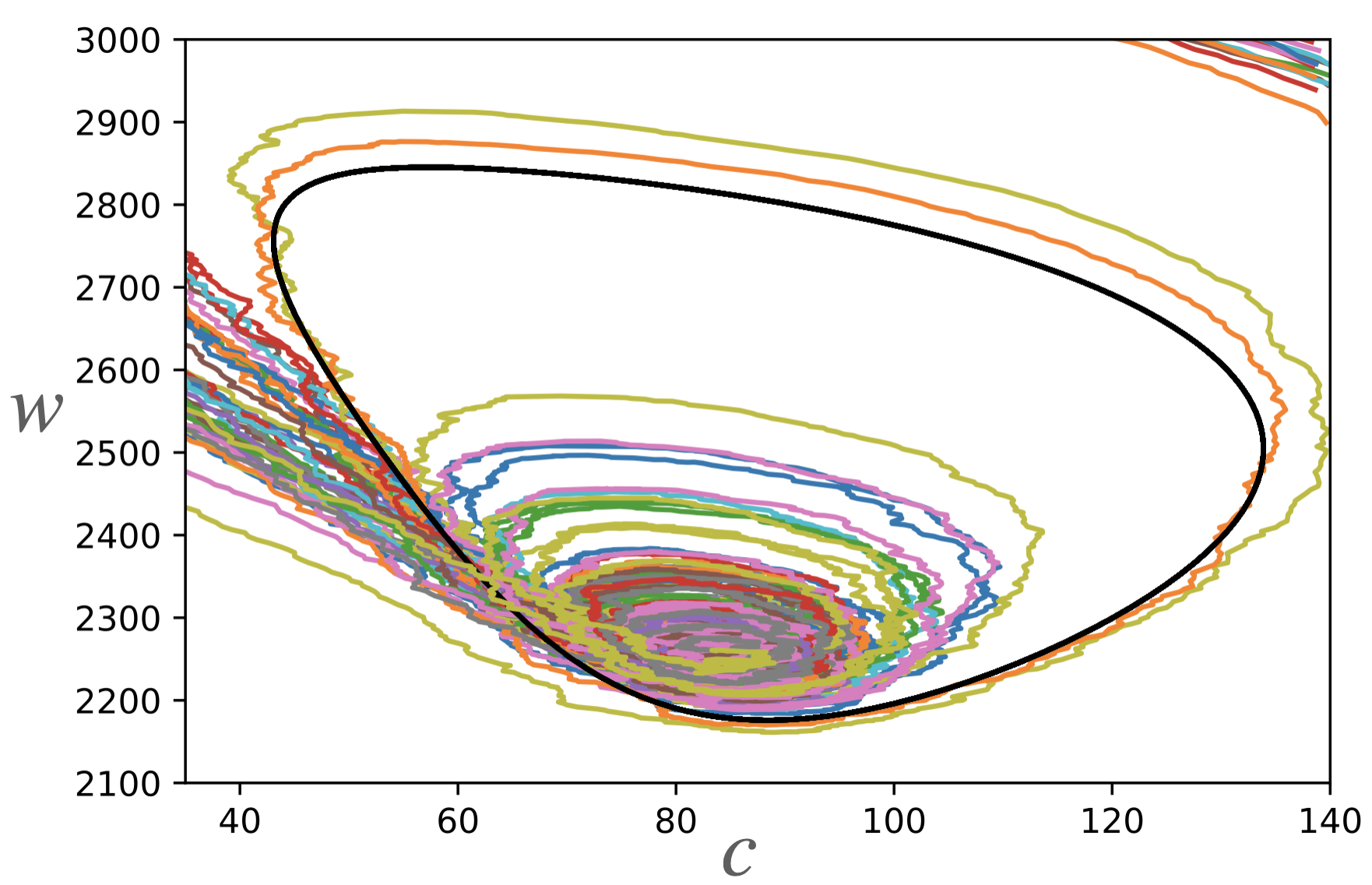}
\caption{Realizations (colored trajectories) of \eqref{EQ:CC stoch} that escape the unstable periodic orbit $\Gamma_u$ (black) on the time interval $[0,15]$ with noise strength $\epsilon=5$. Once they have escaped, they follow the stable periodic orbit $\Gamma_s$ (not pictured) until the end of the time interval.}
\label{Fig:NoisySims}
\end{figure}

The phenomenon of interest is the escape from the stable fixed point to the stable periodic orbit. Using numerically simulated noisy trajectories to gain an initial understanding of the stochastic behavior, there seems to be a clear location on $\Gamma_u$ that the  trajectories concentrate around when tipping away from the stable fixed point, as seen in Figure \ref{Fig:NoisySims}. From Freidlin-Wentzell \cite{freidlin_random_2012} theory, the system will tip with probability one, but it needs to be determined how and on what timescale. The approach is to consider the most probable path for tipping from the stable fixed point to the unstable periodic orbit, as the system will follow the deterministic flow once it reaches the basin boundary.

In the noisy system, for a noise strength in the small noise limit, realizations should cycle around the unstable periodic orbit \citep{day_exit_1996}. However, away from the limit, we see something quite different: the Monte Carlo simulations show there is a specific region on the unstable periodic orbit where trajectories escape. In fact, for small noise away from the limit the cycling behavior is resisted.

\subsection{The most probable path equations}

The theory of large deviations says that the most probable paths of escape between two states, which in this case are a stable fixed point and an unstable periodic orbit, should minimize the Freidlin-Wentzell functional given by
\begin{equation}
    I[\phi]=\frac{1}{2}\int_0^T |\dot{\phi}-P(\phi)|^2 dt.
    \label{EQ:FWfunc}
\end{equation}

Minimizing this functional leads to the Euler-Lagrange equations
\begin{equation}
\begin{aligned}
\ddot{c}&=F_w \dot{w}+FF_c+GG_c-\dot{w}G_c,  \\
\ddot{w}&=G_c \dot{c}+GG_w+FF_w-\dot{c}F_w,
\end{aligned}
\end{equation}
and using a Legendre transform, we can recast the Euler-Lagrange equations as a Hamiltonian system by setting $p=\dot{c}-F$ and $q=\dot{w}-G$. The Euler-Lagrange equations as a Hamiltonian system reads as
\begin{equation}
\begin{aligned}
\dot{c}&=F+p\\
\dot{w}&=G+q \\
\dot{p}&=-F_cp-G_cq \\
\dot{q}&= -F_{w}p-G_{w}q
\end{aligned}
\label{EQ: CC mpep eq}
\end{equation}
where the Hamiltonian is given by
\begin{equation}
H(c,w,p,q)=F(c,w)p+G(c,w)q+\frac{p^2+q^2}{2}.
\label{EQ: CC hamil_func}
\end{equation}

We note a few facts about \eqref{EQ: CC mpep eq}. First, the planes $p=q=0$ are invariant. These invariant subspaces carry the deterministic flow given by system \eqref{EQ: CC short}. Additionally, the fixed point $z^*$ and periodic orbits $\Gamma_u$ and $\Gamma_s$ reappear with their attraction and repulsion reproduced. We slightly abuse notation to allow the same notation of $z^*$, $\Gamma_u$, and $\Gamma_s$ for the fixed point and the unstable and stable periodic orbits, respectively, in reference to both Equations \eqref{EQ: CC short} and \eqref{EQ: CC mpep eq}, as the $p$ and $q$ components are just zero. While the notation stays the same, notice that their stability properties change between the two-dimensional system in \eqref{EQ: CC short} and the four-dimensional system \eqref{EQ: CC mpep eq}. This is the reason for using system \eqref{EQ: CC mpep eq} for determining the most probable paths of escape from $z^*$ out of its basin of attraction.

The stability of $z^*, \Gamma_u$, and $\Gamma_s$ were determined for the two-dimensional system, and now it is necessary to determine their stability in the four-dimensional system. The Jacobian matrix for the linearized flow of \eqref{EQ: CC mpep eq} is given by
\begin{equation}
J(c,w,p,q)=\begin{pmatrix}
    F_c & F_w & 1 & 0 \\
    G_c & G_w & 0 & 1 \\
    -F_{cc}p-G_{cc}q & -F_{cw}p-G_{cw}q & -F_c & -G_c\\
    -F_{wc}p-G_{wc}q & -F_{ww}p-G_{ww}q & -F_w & -G_w
\end{pmatrix}.
\end{equation}

\noindent The eigenvalues of the matrix $J(z^*)$ are given by
\begin{equation}
\begin{aligned}
\lambda_{1,2}=-&0.456437  \pm 3.02052i, \\
\lambda_{3,4}= \hspace{4mm} &0.456437  \pm 3.02052i,
\label{eqn: unstable evecs}
\end{aligned}
\end{equation}
where $\lambda_{3,4}$ are the eigenvalues associated with the unstable eigenvectors that span the two-dimensional unstable subspace. The corresponding eigenvectors are given by $v_{3,4}$. Notice that $\lambda_{1,2}$ are exactly the eigenvalues of \eqref{EQ: Jacobian 2d numbers}, and correspond to the two-dimensional stable subspace of the fixed point $z^*$ in \eqref{EQ: CC mpep eq}, as well as the deterministic flow of \eqref{EQ: CC short}. It follows that the unstable manifold of $z^*$, denoted by $W^u(z^*)$, is two-dimensional. 

$\Gamma_u$ has one stable, one unstable, and two neutral Floquet multipliers. These can be found numerically or determined using the fact that for a Hamiltonian system with two degrees of freedom, there are two neutral Floquet exponents and two that are reciprocals of one another. $\Gamma_u$ has a two-dimensional stable manifold as integrating its one-dimensional stable direction results in a two-dimensional tangent bundle. It follows that $\Gamma_u$ has both a two-dimensional stable manifold, denoted $W^s(\Gamma_u)$, and a two-dimensional unstable manifold in \eqref{EQ: CC mpep eq}, where the two-dimensional unstable subspace aligns with that of the deterministic flow of \eqref{EQ: CC short}. 

Both the unstable manifold of $z^*$ and stable manifold of $\Gamma_u$ lie in $\mathbb{R}^4$, in the complement of the deterministic plane, and will aid in determining the most probable path, as the transverse intersections of $W^u(z^*)$ and $W^s(\Gamma_u)$ are the heteroclinic orbits. Furthermore, any minimizer of the Freidlin-Wentzell functional must be a trajectory of \eqref{EQ: CC mpep eq}, projected on $(c,w)$-space, lying on $W^u(z^*)$.

\subsection{Computing the unstable and stable manifolds}

As we want to find the heteroclinic connection(s) between $z^*$ and $\Gamma_u$, we have to compute $W^u(z^*)$ and $W^s(\Gamma_u)$ respectively. To compute $W^u(z^*)$, we first need to find the local unstable manifold, $W_{loc}^u(z^*)$. Take the fixed point $z^*\in \mathbb{R}^4$ of \eqref{EQ: CC mpep eq} where the linearized system about $z^*$ has eigenvectors $v_{3,4}$ corresponding to the two eigenvalues with positive real part $\lambda_{3,4}$. Initialize using
\begin{equation}
\psi= z^* + \theta_1v_3 + \theta_2 v_4 ,
\end{equation}
where $ |\theta_1|^2+|\theta_2|^2 = r^2$, and $r$ scales how close the initialization is to the fixed point $z^*$. As $v_3 = \bar{v}_4$, it is necessary that $\theta_1 = \bar{\theta}_2$ in order to keep components real by applying the conjugate. Let 
\begin{equation}
\begin{aligned}
 \theta_1 &=r(\cos(x)+i\sin(x))\\
\theta_2 &= r(\cos(x)-i\sin(x))   
\end{aligned}
\end{equation}
where $0 \leq x < 2\pi$. As $x \in [0,2 \pi)$ varies, a collection of $\psi$ values is created, which together form a circle of points that lie in $W_{loc}^u(z^*)$; we denote this set of points $K$. See Figure \ref{fig:CircleKa} for a depiction of $K$.

The global unstable manifold is generated by initiating trajectories from circle $K$ inside the local unstable manifold. System  \eqref{EQ: CC mpep eq} is numerically simulated using each point on $K$ as an initial condition, and the union of the set of these full trajectories form $W^u(z^*)$. With exception of the heteroclinic orbits, the only trajectories of interest are those that reach the unstable periodic orbit $\Gamma_u$ in finite time. Therefore, we need only to integrate over a time period long enough such that a portion of $W^u(z^*)$ reaches just past the torus $\mathcal{T}_{\scriptscriptstyle\Gamma_u}$ on $\Gamma_u \in \mathbb{R}^4$. 
A similar approach was done in \citet{Emman_IVDP} using the parameterization method, an approach well-suited for polynomial vector fields, see \citet{Maciej, FLEU}.

We have to use a different approach to compute $W^s(\Gamma_u)$ as we are calculating the stable manifold of a periodic orbit. The periodic orbit $\Gamma_u$ is discretized into $N+1$ points $\Gamma_{u_k}$, where $0 \leq k \leq N$. For each point, there corresponds a time $\tau_k$ such that $\Gamma_{u_k} =  \Gamma_{u_0}(\tau_k)$. We note that it does not matter the location of the starting point. 

For $k=0$, we find the \textit{monodromy matrix}, the state transition matrix $\Phi$ after one period of $\Gamma_u$. Designate the eigenvector associated with the stable eigenvalue of the monodromy matrix as  $\xi_{s_0}$, where the stable eigenvalue $\lambda \in (0,1)$, and note $\xi_{s_0}$ is tangent to the stable manifold at $\Gamma_{u_0}$. The stability properties of the state transition matrix $\Phi$ remain invariant across all points $\Gamma_{u_k}$ on the periodic orbit, regardless of the specific choice of $k$. Moreover, if the state transition matrix is known at a reference point, such as $\Gamma_{u_0}$, it is possible to compute the eigenvectors of $\Phi$ at any other point $\Gamma_{u_k}$ along the orbit.
Formally this is written as
\begin{equation}
    \xi_{s_k} = \Phi(0, \tau_k)\xi_{s_0}.
\label{EQ: stable mani}
\end{equation}
For each discretized point on $\Gamma_u$, the state transition matrix $\Phi(0, \tau_k)$ is computed to obtain the tangent space to the stable manifold at the point with \eqref{EQ: stable mani}, where the collection of these $ \xi_{s_k}$ form a tangent bundle.

\begin{figure}[]
  \centering
  \subfloat[]{\includegraphics[width=0.35\textwidth]{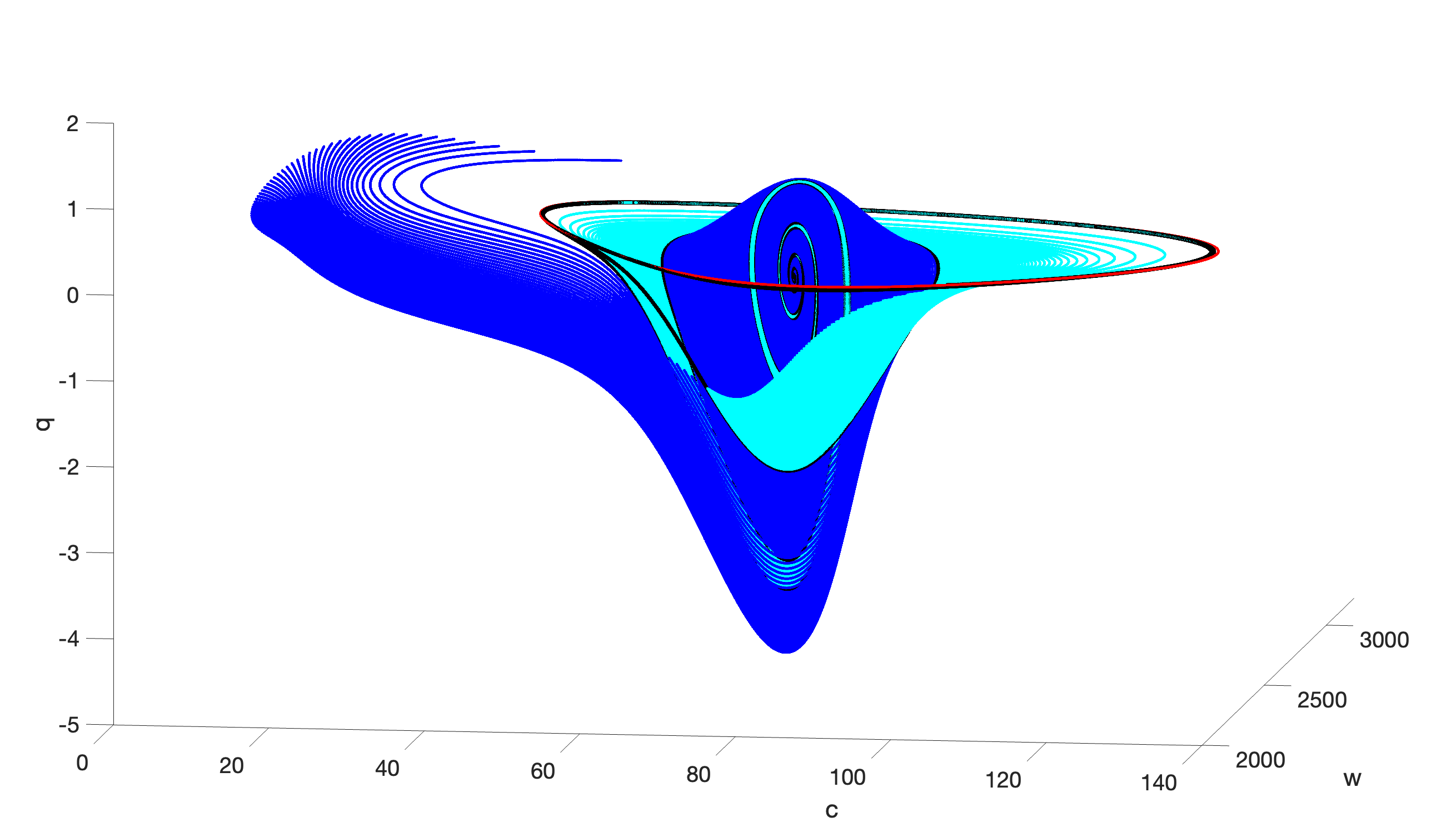}\label{fig:unstable}}
  \hspace{5mm}
  \subfloat[]{\includegraphics[width=0.35\textwidth]{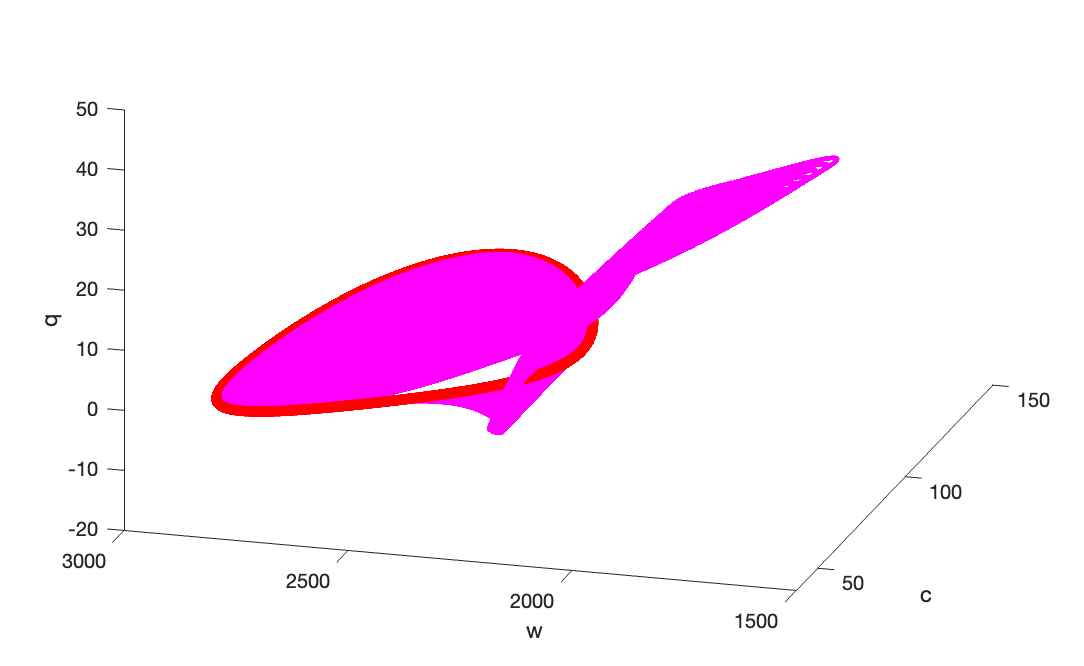}\label{fig:stable}}
  \caption{The invariant manifolds illustrated in $(c,w,q)$ space overlaid with $\Gamma_u$ (red) in the deterministic plane. (a) $W^u(z^*)$. The trajectories of interest (the piece of $W^u(z^*)$ that leaves) are in royal blue. (b) $W^s(\Gamma_u)$. }
   \label{FIG:invariantmanis}
\end{figure}

A tolerance $r$ is chosen, where $r$ scales how close the initialization is to the periodic orbit. For small enough $r$, 
\begin{equation}
    x_{s_k} = \Gamma_{u_k} + r \xi_{s_k}
\label{EQ: ICS stable mani}
\end{equation}
are points in the stable subspace of $\Gamma_u$. The set of initial conditions \eqref{EQ: ICS stable mani} is integrated in backwards, finite time, creating a collection of orbits $x_{s_k}(t)$. The collection of these orbits form the global stable manifold of $\Gamma_u$. See Figure \ref{fig:stable} for a visualization of $W^s(\Gamma_u).$

\subsection{Finding heteroclinic orbits} \label{sec:hets}

Now that there are approximations of $W^u(z^*)$ and $W^s(\Gamma_u)$, it is possible to find the heteroclinic orbits. The heteroclinic orbits are found through the intersections of these invariant manifolds. We expect these intersections to be transverse, which is why they can be found computationally.

To find these intersections, an algorithm to find sets of closest points, one on $W^u(z^*)$ and the other on $W^s(\Gamma_u)$, is used. The algorithm for finding a set of closest points is as follows.
\begin{enumerate}
\item Restructure both the unstable and stable manifold datasets into two dataframes with columns of $c,w,p,q$.
\item Create a BallTree for the stable manifold, which partitions its dataframe into a nested set of balls. The resulting data structure has characteristics that make it useful for a nearest neighbor search.
\item Query the tree to find its nearest neighbor within the unstable manifold dataframe, and return the distance between, and the IDs, of the two points. Sort by the shortest distance.
\end{enumerate}

Initially the heteroclinic orbits are computed using these sets of points, where the point on $W^s(\Gamma_u)$ is integrated forward in time towards $\Gamma_u$, and the point on $W^u(z^*)$ is integrated backwards in time towards $z^*$. The tangency of the two trajectories has to be verified to ensure that the two points actually have trajectories tangent to one another (not intersecting one another or meeting at a cusp). This process is performed by inspection of the integration near the intersection point. Once it is determined that they fall on the same path, the trajectory integrated backwards in time is used to find its corresponding point on the parameterized circle $K$.  Using this process, two heteroclinic orbits are found: $\mathcal{H}_1$ and $\mathcal{H}_2$. Therefore there are two points on $K$ corresponding to two angles $x_1,x_2 \in [0,2 \pi)$ that give us  $\mathcal{H}_1$ and $\mathcal{H}_2$.

\begin{figure*}[ht]
  \centering
  \subfloat[]{\includegraphics[width=0.3\textwidth]{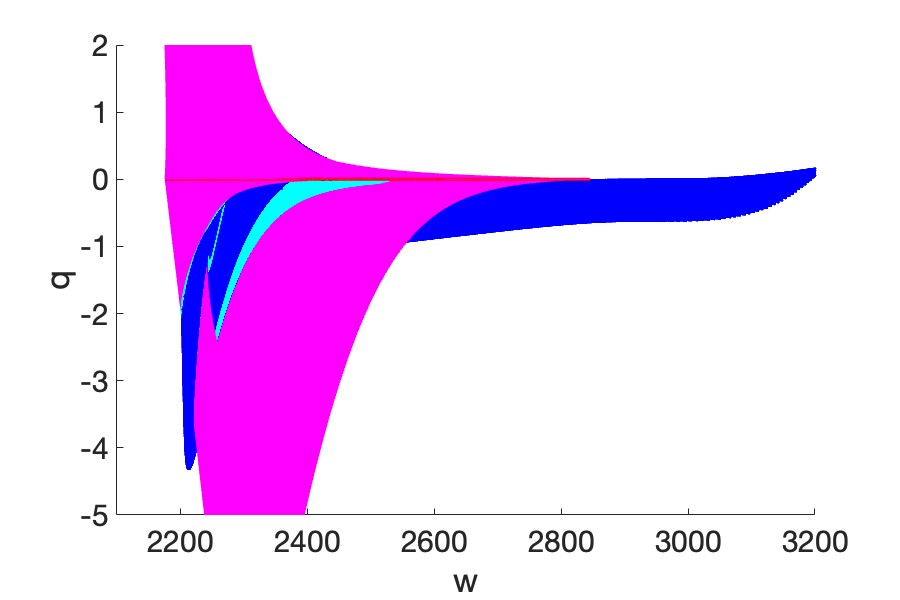}}
  \hspace{5mm}
  \subfloat[]{\includegraphics[width=0.3\textwidth]{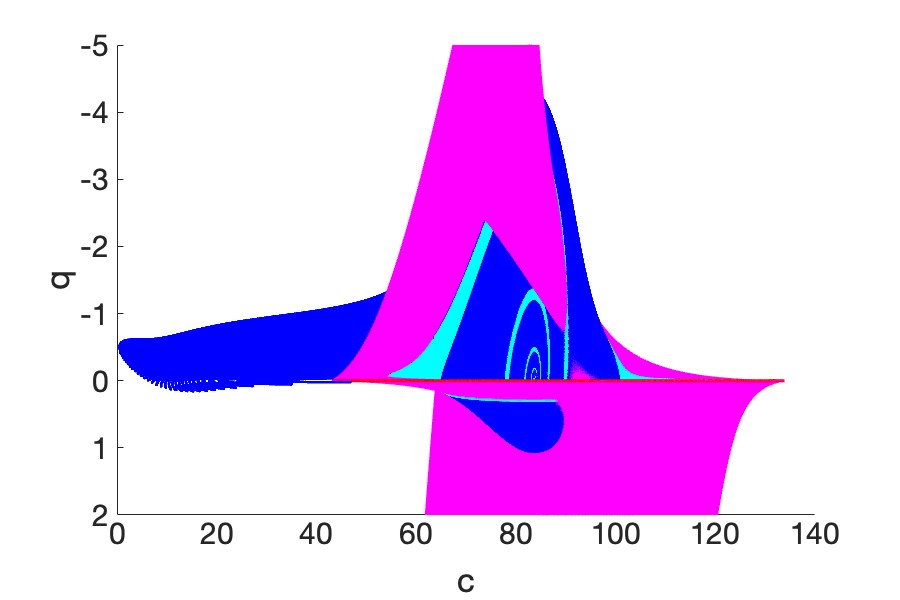}}
   \hspace{5mm}
    \subfloat[]{\includegraphics[width=0.3\textwidth]{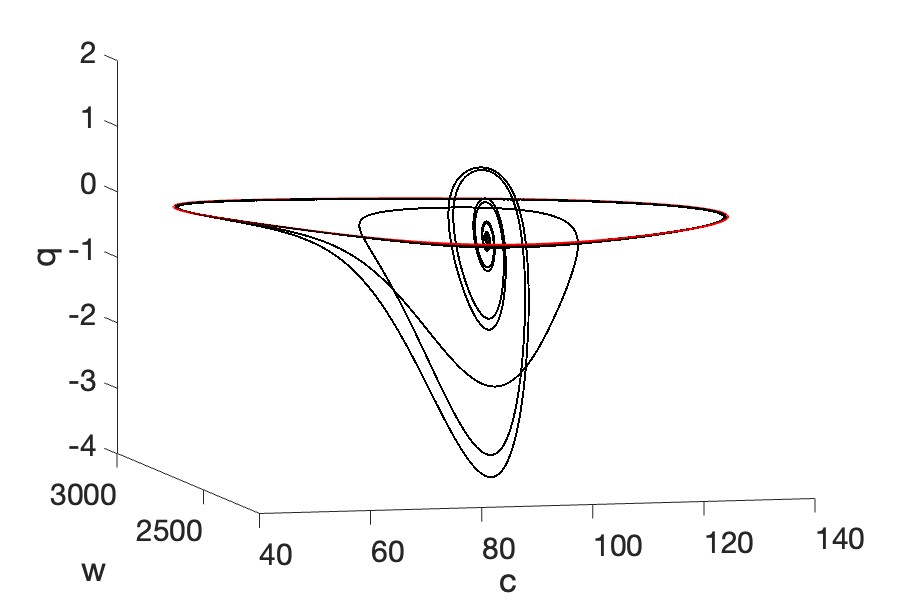}}
  \caption{The invariant manifolds illustrated in $(c,w,q)$ space overlaid with $\Gamma_u$ (red) in the deterministic plane. In both (a) and (b), we have $W^u(z^*)$ (blue) and $W^s(\Gamma_u)$ (magenta) displayed together. Their transverse intersections (black) correspond to the heteroclinic orbits, depicted in (c).}
   \label{FIG:invariantmanis2}
\end{figure*}

An overlap of the invariant manifolds with their transverse intersections are illustrated in $(c,w,q)$ space in Figure \ref{FIG:invariantmanis2}. 

\subsection{The River} \label{Sec: River}
We found that there are exactly two heteroclinic orbits, $\mathcal{H}_1$ and $\mathcal{H}_2$, connecting $z^*$ and $\Gamma_u$. As mentioned in Section \ref{sec:hets}, the heteroclinic orbits $\mathcal{H}_1$ and $\mathcal{H}_2$ are associated with two points on $K$, which correspond to two angles $x_1$ and $x_2$ respectively. We use the natural angle ordering on $K$ and see that then $x_1 < x_2$. The angles $x_1$ and $x_2$ divide the circle $K \subset W^u(z^*)$ into two arcs of points. From Lemma 2.3 in \citet{Emman_IVDP}, there are angles $x$ near angle $x_1$ for which the associated trajectories pass over $\Gamma_u$, and so they are considered exit trajectories. These exit trajectories correspond to $x$ values between $x_1$ and $x_2$, and, in a measure-theoretic sense on $K$ most exit with minimal rotation.

It can be seen that all trajectories with angles between $x_1$ and $x_2$ exit $\Gamma_u$. This implies that every trajectory associated with angles $x \in (x_1, x_2)$ crosses $\Gamma_u$ when projected into $(c,w)$ space. We use the term  \textit{River} to refer to this set of trajectories on $W^u(z^*)$ that cross $\Gamma_u$. Let $Z$ be a trajectory $z(t)$ for $t \in (-\infty,t_f]$ satisfying \eqref{EQ: CC mpep eq}. Formally, the river, denoted $\mathcal{R}$, is defined as
{\small
\[\mathcal{R}= \{Z \hspace{1mm} | \hspace{1mm} z (\tau) \in K(x),\ \text{for some }x_1< x< x_2,\ \tau <t_f,\ z(t_f) \subset \mathcal{T}_{\scriptscriptstyle\Gamma_u}\}\]
}

The heteroclinic orbits $\mathcal{H}_1$ and $\mathcal{H}_2$ form the "banks" of the river. The Maslov Index is equal to the number of conjugate points along a trajectory, see \cite{Milnor} for a definition of conjugate point. Formally, $M(Z)$ is the Maslov Index of the trajectory $Z$, which is defined as the number of conjugate points along $z(t)$ for $t \in (-\infty,t_f]$ (where $t_f=\infty$ is allowed, counting multiplicity). The two heteroclinic orbits are distinguished by their Masolv Index: $M(\mathcal{H}_1)=0$ and $M(\mathcal{H}_2)=1$.

Using the Maslov Index, a subset of the river can be further identified, called the subriver.  Let $Z$ be a trajectory in $\mathcal{R}$. The sub-river, denoted $\tilde{\mathcal{R}}$, is defined as
\[\tilde{\mathcal{R}}= \{Z \in \mathcal{R}| M(Z)=0\},\] where $M(Z)$ is the Maslov Index of the trajectory $Z$. $M(Z)$ in $\mathcal{R}$ on $W^u(z^*)$ is the number of conjugate points along $z(t)$ for $t \in (-\infty,t_f]$, counting multiplicity. 

\begin{figure}[ht]
  \centering
  \subfloat[]{\includegraphics[width=0.23\textwidth]{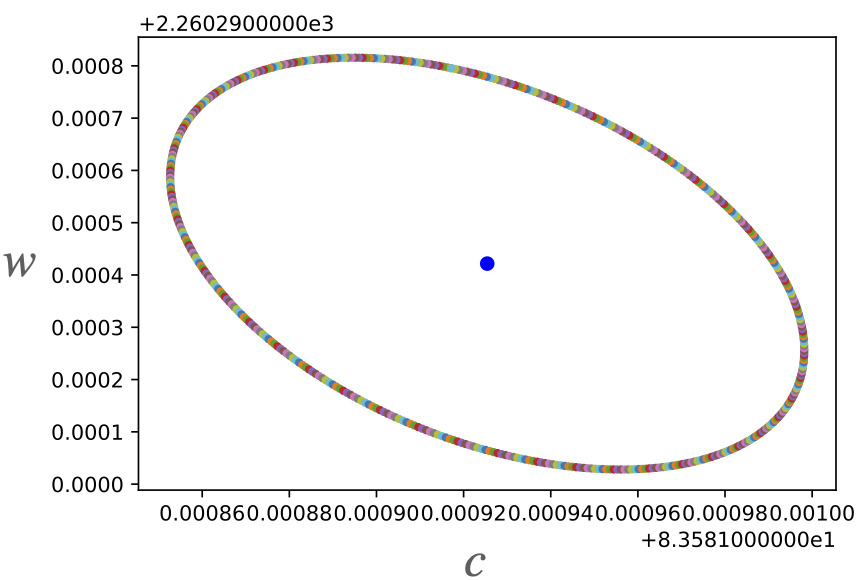}\label{fig:CircleKa}} 
  \hspace{1mm}
  \subfloat[]{\includegraphics[width=0.23\textwidth]{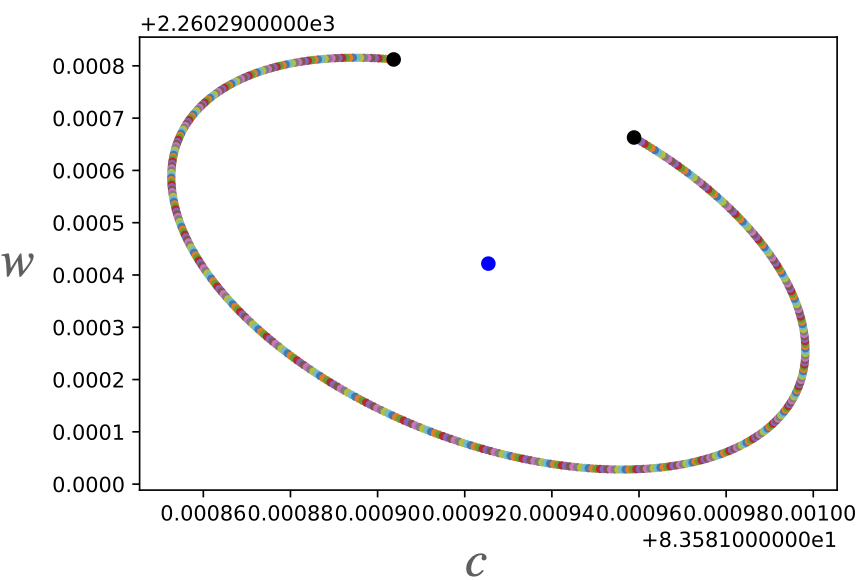}\label{fig:CircleKb}} 
  \subfloat[]{\includegraphics[width=0.23\textwidth]{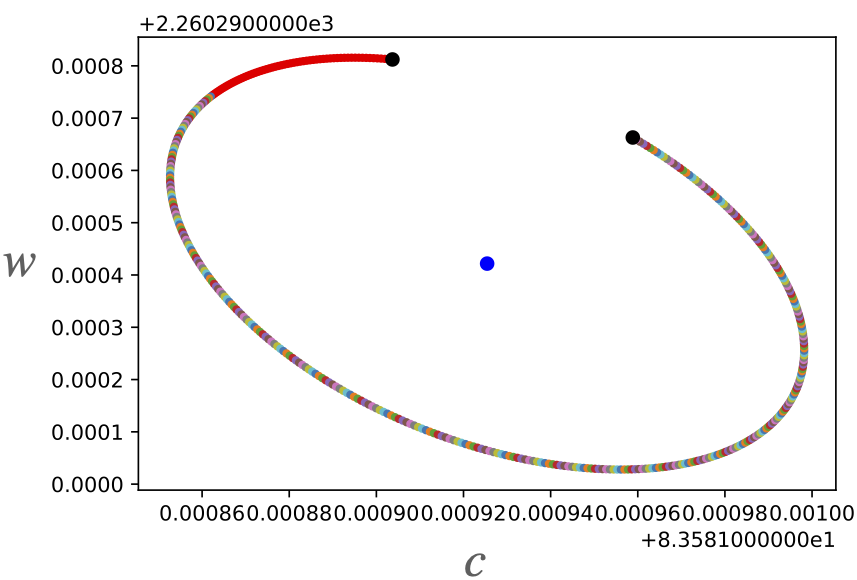}\label{fig:CircleKc}} 
  \caption{The stable fixed point $z^*$ is denoted by the blue circle. The black circles represent the angles $x_1$ and $x_2$ associated with the heteroclinic orbits $\mathcal{H}_1$ and $\mathcal{H}_2$. (a) The collection of $\psi$ values that form the set $K$. This circle lies in $W^u_{loc}(z^*)$ and integrated forward in time to find $W^u(z^*)$. (b) The arc of set $K$ that will form the river trajectories when integrated forward in time. (c) The arc of set $K$ that will form the river trajectories overlaid with the angles that will form the subriver (red).}
   \label{FIG:CircleK}
\end{figure}

We make the assumption that the subriver is a connected set. This is easily verified in our problem, but may not be true in general.  In our problem, the sets of trajectories originating from the circle $K$, which forms the river, partition into two connected components: one with Maslov Index $0$ and the other with Maslov Index $1$. This partition is illustrated by the connected component highlighted in red on $K$ in Figure~\ref{fig:CircleKc}.

Let $\Phi: K(x_1, x_2) \to \mathcal{T}_{\scriptscriptstyle\Gamma_u}$ be a transition map. Define 
\[\hat{x} = \inf\{x \mid \Phi(K(x)) \in \tilde{\mathcal{R}}\}.\]
We will refer to $\Phi(K(\hat{x}))$ as the {\it pivot point}. The trajectory associated with the pivot point will have Maslov Index 1 and the conjugate point will occur as the trajectory crosses $\mathcal{T}_{\scriptscriptstyle\Gamma_u}$, which marks the transition between trajectories in the sub-river $\tilde{\mathcal{R}}$ and those outside it. 

Many recent results use the Maslov Index to prove that one can count unstable eigenvalues by counting conjugate points, see for example \citet{beck_validated_2022, Beck}. This also implies instability relative to the gradient flow associated with the action functional (Morse Theory, see \citet{Milnor}). We note that $\tilde{\mathcal{R}}\subset \mathcal{R}$.

\begin{remark}
    It is important to emphasize that the river $\mathcal{R}$ and its subset, the sub-river $\tilde{\mathcal{R}}$, are collections of trajectories, not exit points. 
\end{remark}

\begin{figure}[ht]
\subfloat[]{\includegraphics[width=0.3\textwidth]{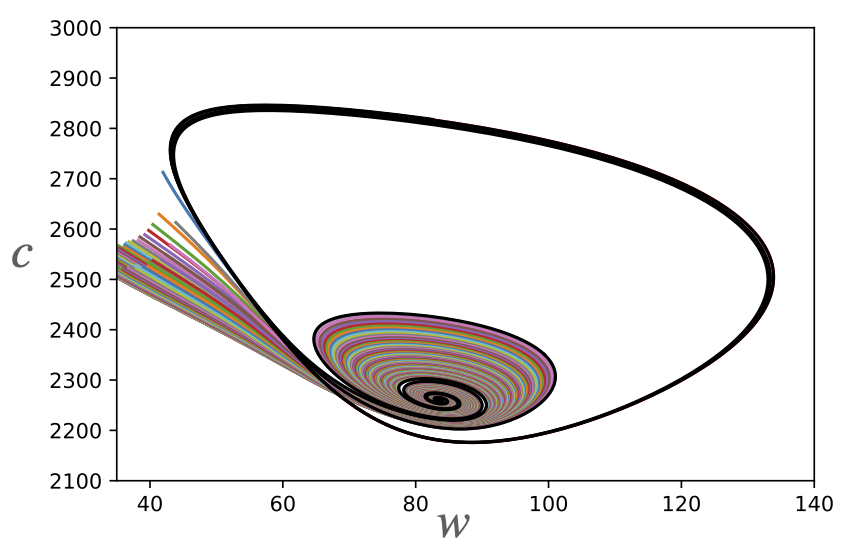}\label{fig:river2d}}
\hspace{1mm}
\subfloat[]{\includegraphics[width=0.25\textwidth]{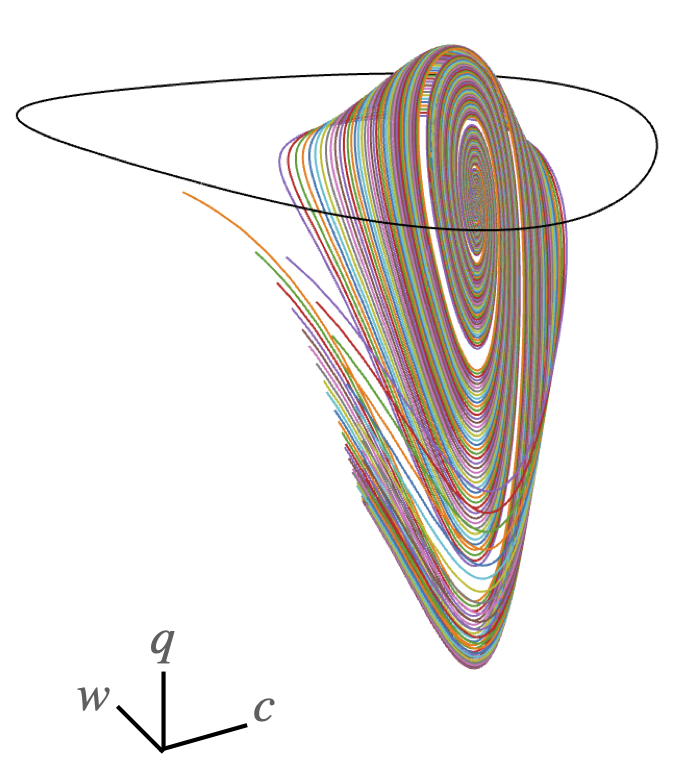}\label{fig:river3d}}
\caption{(a) The River trajectories (colored curves) integrated until they are past $\Gamma_u$. The heteroclinic orbits $\mathcal{H}_1$ and $\mathcal{H}_2$ (both in black) form the banks of the River. (b) The River integrated until $\mathcal{T}_{\scriptscriptstyle\Gamma_u}$ illustrated in $(c,w,q)$ space.}
\label{fig:Riverall}
\end{figure}

The points on $K$ that correspond to $\mathcal{R}$ are depicted in Figure \ref{fig:CircleKb} and the points on $K$ that correspond to $\tilde{\mathcal{R}}$ are shown in red in Figure \ref{fig:CircleKc}. $\mathcal{R}$ with both $\mathcal{H}_1,\mathcal{H}_2$ is depicted in two dimensions in Figure \ref{fig:river2d}, while the $\mathcal{R}$ itself is shown in three dimensions in Figure \ref{fig:river3d}.

The definitions of the River, subriver, and the Maslov Index, imply it is only necessary to find the number of conjugate points on a River trajectory to determine whether or not it is a trajectory in the subriver. A conjugate point occurs along a trajectory $Z$ in $W^u(z^*)$ when the tangent space to the invariant manifold at a point on $Z$ has a degenerate projection onto $(c,w)$ space \cite{Emman_IVDP}. These points can be found by tracking the tangent space to $W^u(z^*)$ along trajectories in $W^u(z^*)$. There are coordinates on the space of planes that allow the tracking of this tangent space to $W^u(z^*)$ along a trajectory. The approach is to form the Pl{\"u}cker coordinates of the two-dimensional subspace in $\mathbb{R}^4$, as Pl{\"u}cker coordinates are a way to assign six homogeneous coordinates to planes in $\mathbb{R}^4$. The space of planes in $\mathbb{R}^4$ can be embedded into $\mathbb{P}^5$.

A conjugate point can be conveniently described in Pl{\"u}cker coordinates. Using Lemma 4.3 from \citet{Emman_IVDP}, the time $t=\tau$ is a conjugate point for a trajectory $z(t)=(c(t),w(t),p(t),q(t))$ in $W^u(z^*)$ if $\rho_{12}(\tau)=0$ for the Pl{\"u}cker coordinates of the tangent space of $W^u(z^*)$. We will now describe the derivation of the Pl{\"u}cker formulation for our problem.

If $\Pi$ is a two-dimensional subspace of $\mathbb{R}^4$, using coordinates $(y_1,y_2,y_3,y_4)$, then the coordinates of $\Pi$ are given by 
\begin{equation}
 dy_{i_1} \wedge dy_{i_2}
 \label{eq:wedge}
\end{equation}
for all choices of $(i_1,i_2)$ \cite{Jones1995}. Let $\Pi$ be a plane spanned by $v_3$ and $v_4$ from \eqref{eqn: unstable evecs} with
\begin{equation}
v_3=\begin{pmatrix}v_{31}\\v_{32}\\v_{33}\\v_{34} \end{pmatrix} \textrm{and }
v_4=\begin{pmatrix}v_{41}\\v_{42}\\v_{43}\\v_{44}
\end{pmatrix}.
\end{equation}

\noindent Use \eqref{eq:wedge}, to set
    \begin{equation}
    \rho_{ij} = \begin{vmatrix}
    v_{3i} & v_{3j}\\
    v_{4i} & v_{4j}
    \end{vmatrix} = dx_i \wedge dx_j(v_3,v_4), \quad 1\leq i,j \leq 4, \quad i \neq j.
    \end{equation}
It follows the Pl{\"u}cker coordinate of interest for this particular problem is
\begin{equation}
\begin{aligned}
\rho_{12}&=dc \wedge dw.
\end{aligned}
\end{equation}

The variation of a plane's Pl{\"u}cker coordinates can be captured by an ordinary differential equation looking at the coordinates variation in time. This differential equation is calculated using differential forms of the linearized system \eqref{EQ: CC mpep eq}. 
This new system inherits a Hamiltonian structure with allows us to connect it's evolution to the space of Lagrangian planes $\Lambda(2)$. This allows one to define a phase in $\Lambda(2)$ and the standard definition of the Maslov Index is that it counts the winding of this phase. The system of interest is:

\begin{equation}
\dot{U} = BU,
\label{EQ: pluck DE}
\end{equation}
where $U=(\rho_{12},\rho_{13},\rho_{14},\rho_{23},\rho_{24},\rho_{34})$,  
\begin{widetext}
\begin{equation}
B(c,w,p,q) = \begin{pmatrix}
F_c+G_w& 0&1 &-1&0&0\\
-pF_{cw}-qG_{cw}&0&-G_c&F_w&0&0\\
-pF_{ww}-qG_{ww}&-F_w&F_c-G_w&0&F_w&1\\
pF_{cc}+qG_{cc}&G_c&0&-F_c+G_w&-G_c&-1\\
pF_{wc}+qG_{wc}&0&G_c&-F_w&0&0\\
0&pF_{cw}+qG_{cw}&-pF_{cc}-qG_{cc}&pF_{ww}+qG_{ww}&-pF_{cw}-qG_{cw}&-F_c-G_w
\end{pmatrix},
\end{equation}
\end{widetext}
and $\rho_{13}=dc \wedge dp$, $\rho_{14}=dc \wedge dq$, $\rho_{23}=dw \wedge dp$, $\rho_{24}=dw \wedge dq$, $\rho_{34}=dp \wedge dq$.

For each trajectory in $\mathcal{R}$, $z(t)$, initialized from $K$, the locations $(c,w,p,q)$, as well as the time of each point, are known. We form Pl{\"u}cker coordinates of the unstable subspace of system \eqref{EQ: CC mpep eq} at $z^*$ and initialize system \eqref{EQ: pluck DE} with these coordinates. 
System \eqref{EQ: pluck DE} is integrated forward in time, by one time step. The ending location becomes the next initial condition, and this process is repeated until reaching the final time of $z(t)$ (e.g. reaching $\mathcal{T}_{\scriptscriptstyle\Gamma_u}$). Throughout this process we record $\rho_{12}$, and then ultimately find the values of $t$ where $\rho_{12}=0$.

\begin{figure*}[ht]
  \centering
  \subfloat[]{\includegraphics[width=0.3\textwidth]{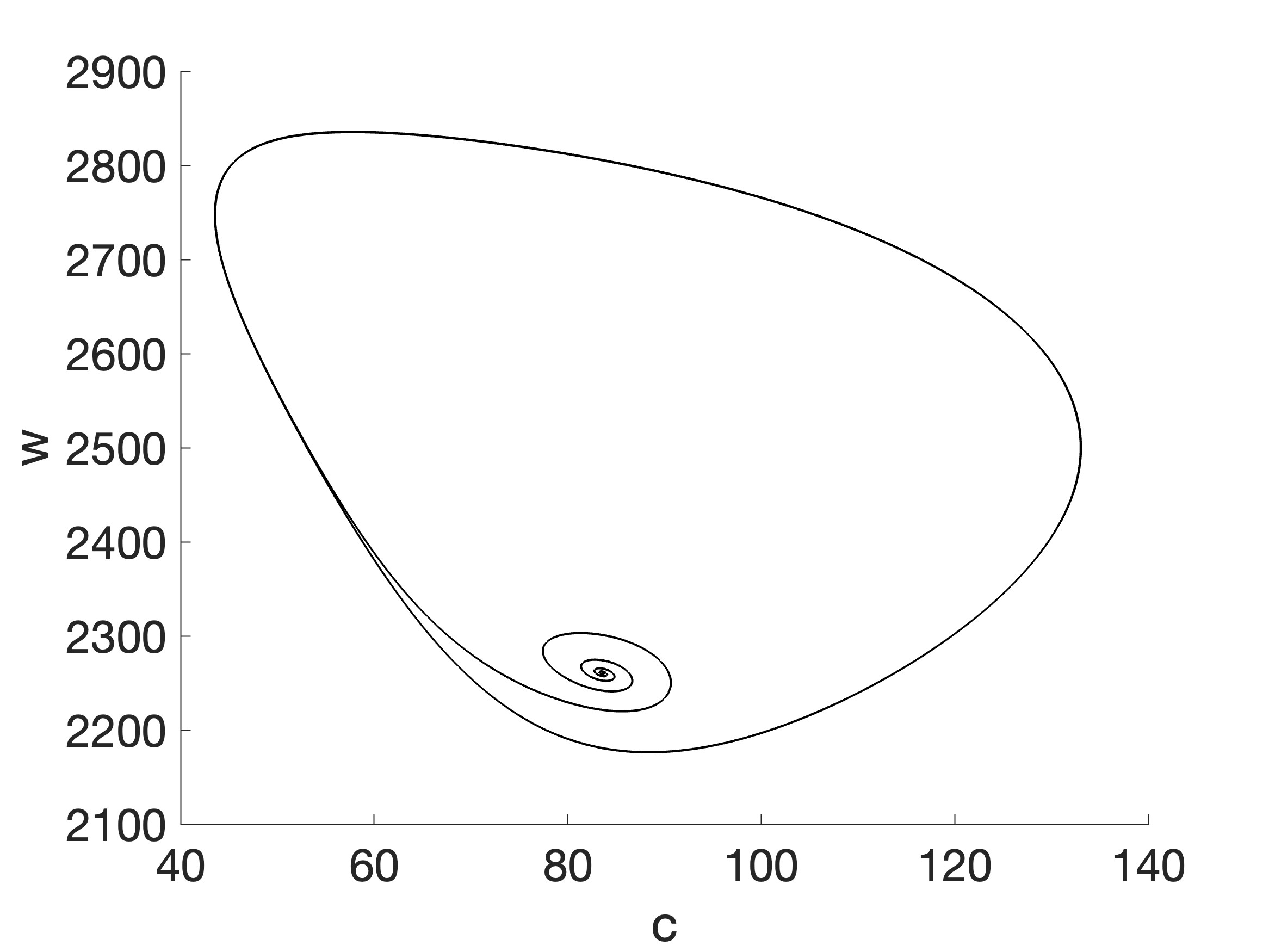}} 
  \hspace{10mm}
  \subfloat[]{\includegraphics[width=0.3\textwidth]{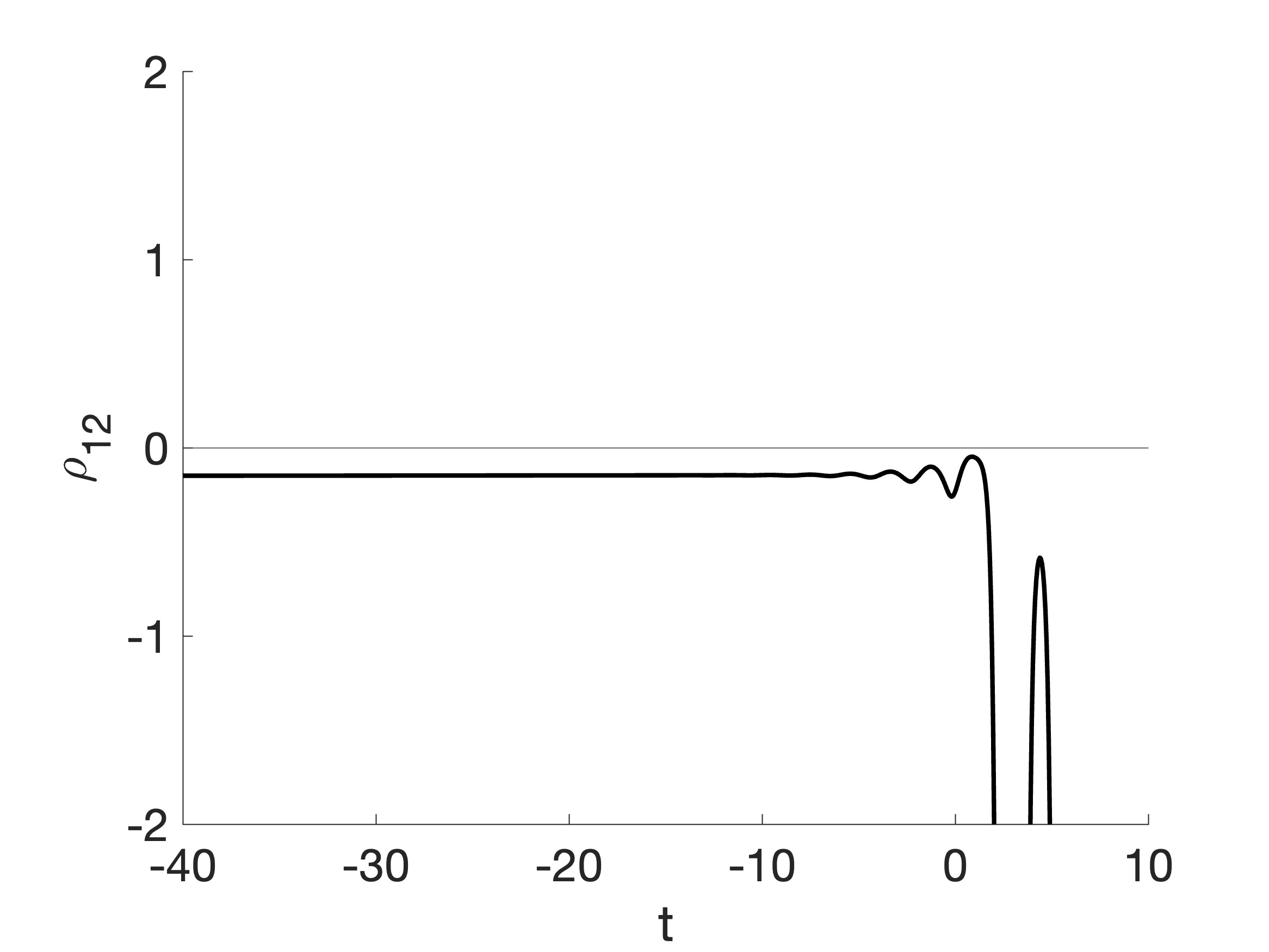}\label{fig:mas1}} 
   \hspace{10mm}
   \\
  \subfloat[]
  {\includegraphics[width=0.3\textwidth]{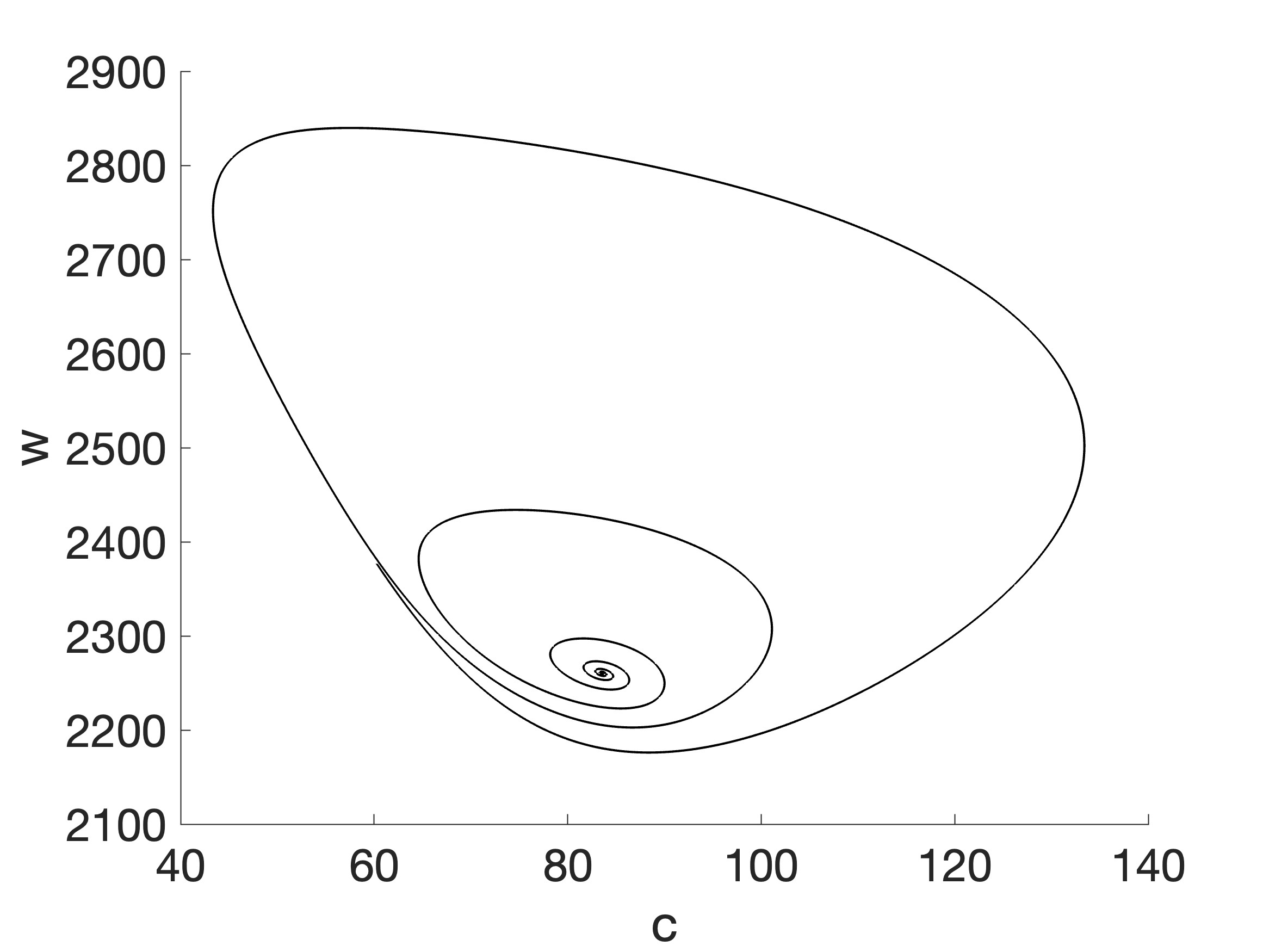}} 
  \hspace{10mm}
  \subfloat[]{\includegraphics[width=0.3\textwidth]{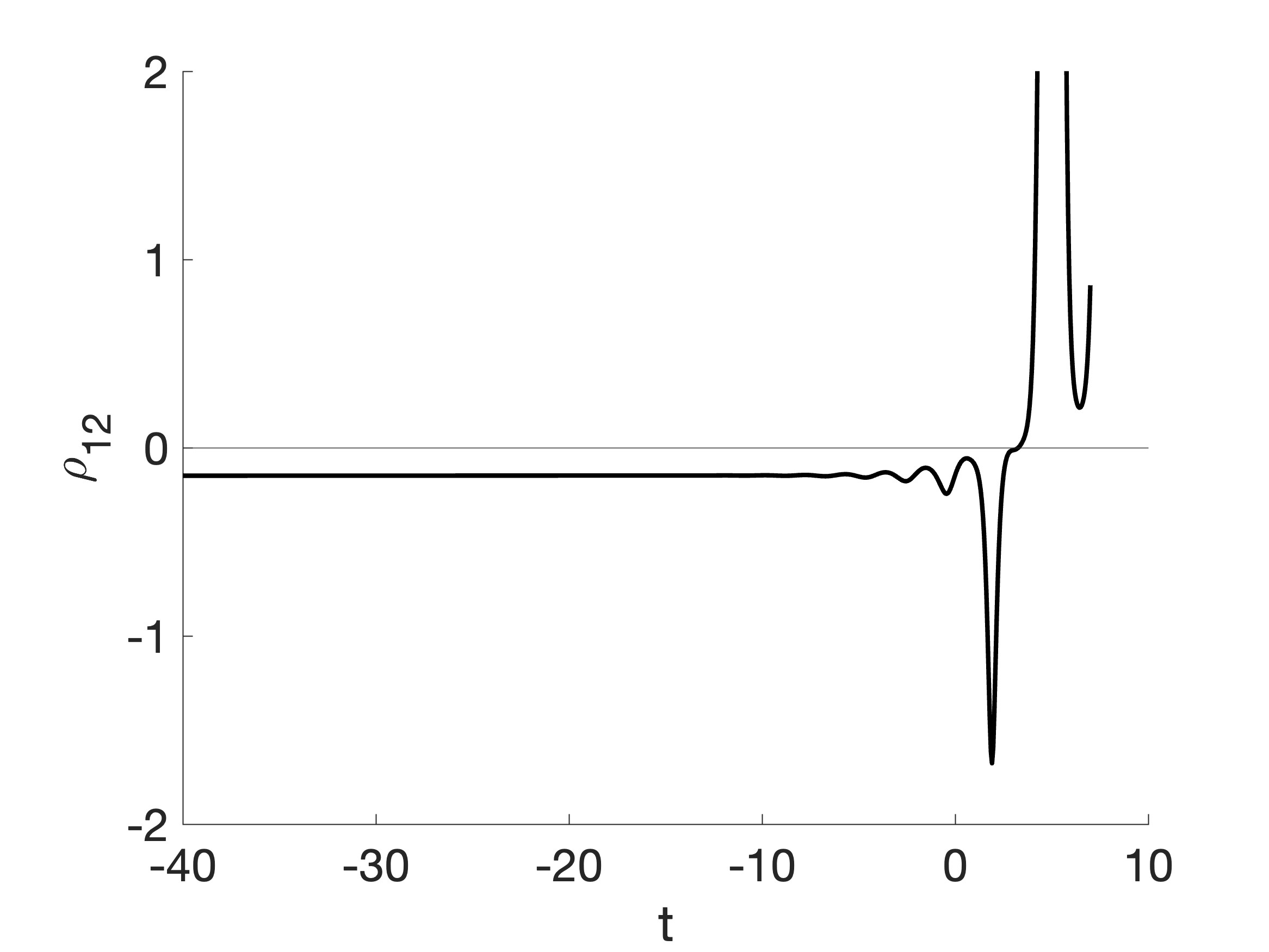}\label{fig:mas2}} 
\caption{(a) The heteroclinic orbit $\mathcal{H}_1$ of system \eqref{EQ: CC mpep eq}. (b) A plot tracking the component $\rho_{12}$ for $\mathcal{H}_1$ and showing there are no conjugate points on the time interval specified as the curve never crosses 0. (c) The heteroclinic orbit $\mathcal{H}_2$ of system \eqref{EQ: CC mpep eq}. (d) A plot tracking the component $\rho_{12}$ for $\mathcal{H}_2$ and showing there is one conjugate point on the time interval specified as the curve crosses 0 once.
}
\label{fig:Hetero-Maslov}
\end{figure*}

Figure \ref{fig:Hetero-Maslov} illustrates the tracking of conjugate points for the heteroclinic orbits $\mathcal{H}_1$ and $\mathcal{H}_2$ that were computed in Section \ref{sec:hets}. The heteroclinic orbits are shown in the left column and their associated plot in the right column indicates when a conjugate point is detected. These computations show that one heteroclinic orbit has Maslov Index 0 (Figure \ref{fig:mas1}) and the other has Maslov Index 1 (Figure \ref{fig:mas2}). The zero-conjugate point heteroclinic orbit is denoted by $\mathcal{H}_1$ and the one-conjugate point heteroclinic orbit is denoted by $\mathcal{H}_2$.

Using Theorem 5.2 in \citet{Emman_IVDP}, the trajectories in $\tilde{\mathcal{R}}$ are local minimizers of the Freidlin-Wentzell functional given their respective boundary values. 
They also play a fundamental role for the noisy realizations that escape for small but not vanishing noise strengths. In the case of the vanishingly small noise limit, $\mathcal{H}_1$ is the most probable escape path.

\begin{remark}
It is crucial to note that the set $\tilde{\mathcal{R}} \cap \mathcal{T}_{\scriptscriptstyle\Gamma_u}$, when projected onto $\Gamma_u$, covers the periodic orbit multiple times. This multi-covering property underscores the complexity of the escape dynamics and highlights why the sub-river alone is insufficient to pinpoint the most probable exit site.
\end{remark}

Indeed let $\mathcal{Q} = \Phi(K(x_1, \hat{x}))$. It has the following properties:
\begin{enumerate}
    \item $\mathcal{Q}$ is an infinite spiral in $\mathcal{T}_{\scriptscriptstyle\Gamma_u}$.
    \item Its projection onto the $(c,w)$-space is all of $\Gamma_u$.
    \item It is pinned at one end by the pivot point.
    \item Every point in $\mathcal{Q}$, except the pivot point, is a local minimizer of the FW action functional with fixed boundary conditions.
\end{enumerate}

 The set $\mathcal{Q}$ plays a key role in determining the escape hatch. But it is still too large. The biggest issue lies behind the cycling phenomenon, namely that there is no global minimizer of the FW functional which crosses $\Gamma_u$. 

\subsection{Monte Carlo simulations and escaping paths}
Our goal was to find a most probable path of escape through an unstable periodic orbit that forms the boundary of the basin of attraction of a stable fixed point. We show that for small noise away from the limit, noisy trajectories that escape will not exhibit infinite cycling behavior but actually escape at a specific region of the periodic orbit. 

We show that this escape region exists by carrying out Monte Carlo simulations on the oceanic carbon cycle model with additive noise. In the computations, the noise strength is set at $\epsilon=5$. The noise strength was chosen such that there was less than 5\% of realizations that escape while also maintaining a noise strength for which numerical convergence was possible for the exit distributions. We consider \eqref{EQ: CC EM} with $\epsilon=5$ on the time interval $[0,15]$, initialized at $z^*$ with a step size of $dt=.005$.

System \eqref{EQ:CC stoch} is numerically simulated using the Euler-Maruyama method, which creates a discretized Markov process \cite{higham_algorithmic_2001}, over the time interval $[0,15].$ The time interval is partitioned into sub-intervals of width $\Delta t=.005$, and the solution is initialized at $c=c^*$ and $w=w^*$. To create the discretized Markov process, the system is recursively defined as

\begin{equation}
\begin{aligned}
c_{n+1}&=(f(c_n)[\mu(1-bs(c_n,c_p)-\theta \bar{s}(c_n,c_x)-\nu)+w_n-w_0])\Delta t \\
&+ \epsilon \Delta W_{1n}, \\
w_{n+1}&=( \mu[1-bs(c_n,c_p)+\theta \bar{s}(c_n,c_x)+\nu]-w_n+w_0) \Delta t \\
&+ \epsilon \Delta W_{2n}.
\end{aligned}  
\label{EQ: CC EM}
\end{equation}

\begin{figure}[ht]
  \centering
  \subfloat[]{\includegraphics[width=0.22\textwidth]{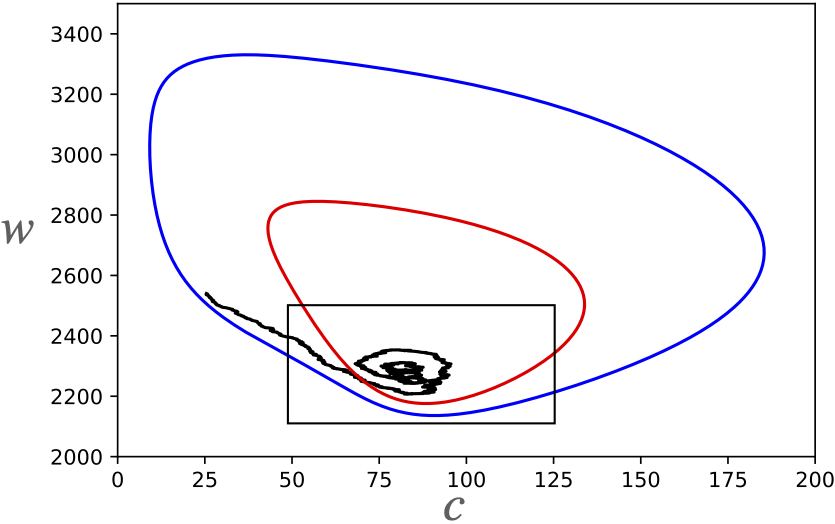}}
  \subfloat[]{\includegraphics[width=0.22\textwidth]{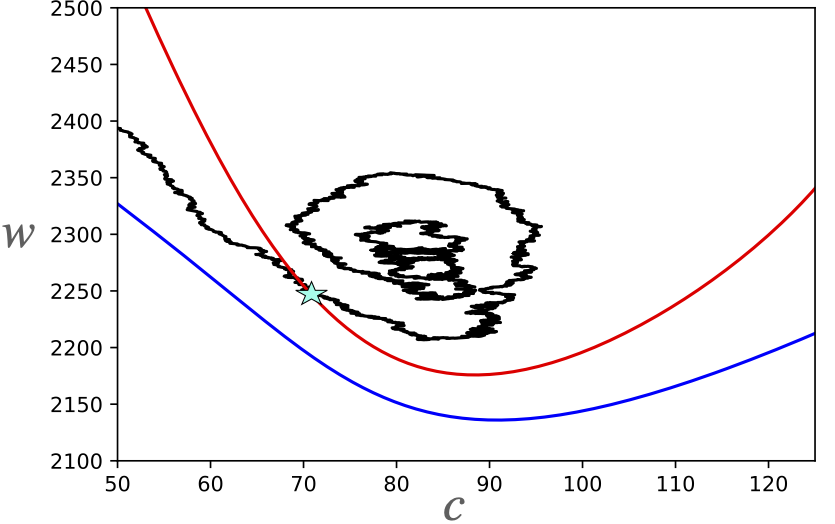}}
\caption{A sample path of \eqref{EQ: CC EM} (black) initialized at $z^*$ on the interval $[0,15]$ with $dt=.005$ and $\epsilon=5$, overlaid with the stable (blue) and unstable (red) periodic orbits. (a) The sample path escapes the unstable periodic orbit, $\Gamma_u$. (b) A zoomed-in version of the rectangular box in (b), where the cyan star denotes the path's exit location, $(c_i,w_i)$.}
\label{fig:tipnotipCC}
\end{figure}

The goal is to find the realizations that have transitioned from $z^*$ to somewhere outside the unstable periodic orbit, and capture where on $\Gamma_u$ they have exited. Let $\tau_i$ denote the first time a path, $X_i$, crosses $\Gamma_u$. Escape events are defined to be the paths $X_i$ that have $\tau_i \leq 15$. Let the point of $X_i$ at $\tau_i$ be given by $(c_i,w_i)$. Refer to Figure \ref{fig:tipnotipCC} for an example of a realization that has escaped on the finite time interval, and the escape location. Assume for $M$ realizations there are $\mathcal{K}$ escape events. We construct the distributions for the $c$ and $w$ locations for the $\mathcal{K}$ escape events. To verify these distributions have converged, the convergence process from Section 7.2 of \citet{Emman_IVDP} is used for both $c$ and $w$ escape locations. 

\begin{figure*}[ht]
  \centering
  \subfloat[]{\includegraphics[width=0.3\textwidth]{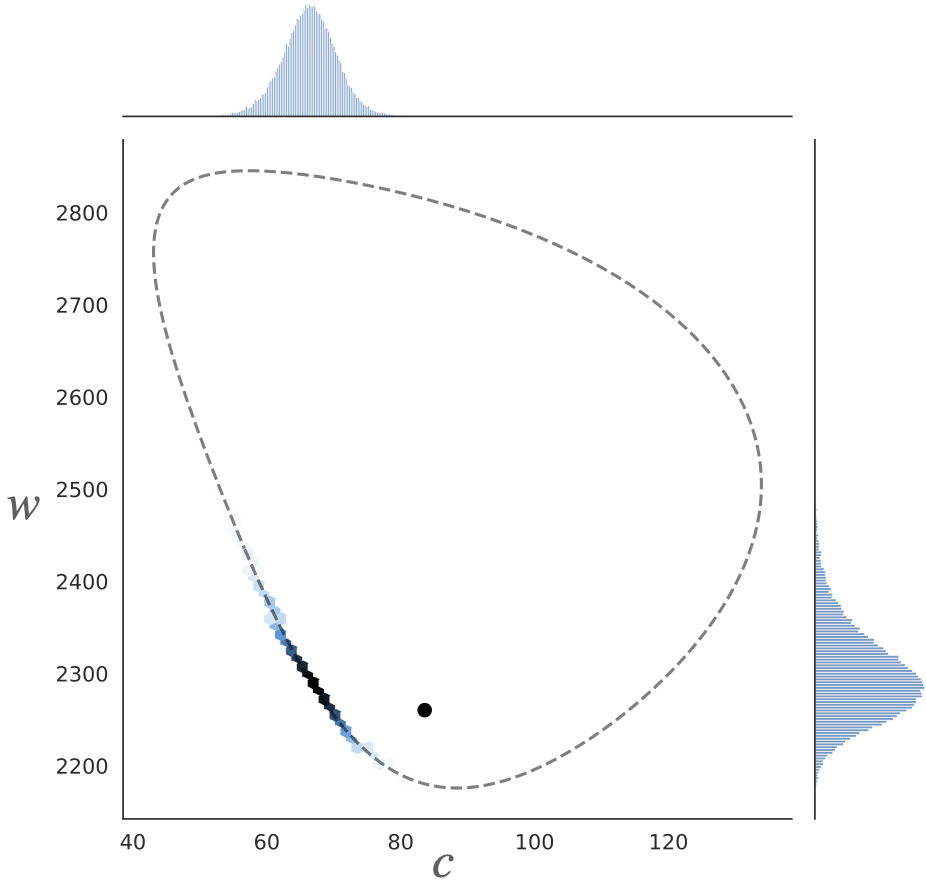}\label{FIG:KDEsa}}
  \hspace{1mm}
  \subfloat[]{\includegraphics[width=0.3\textwidth]{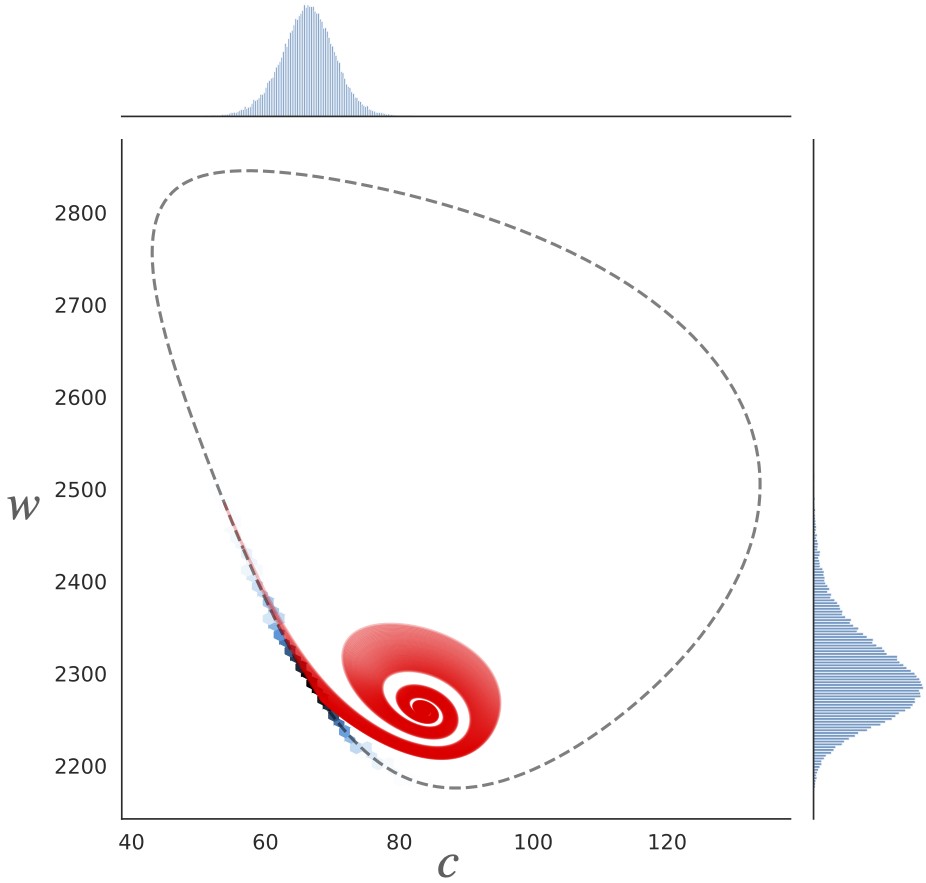}\label{FIG:KDEsb}}
   \hspace{1mm}
  \subfloat[]{\includegraphics[width=0.3\textwidth]{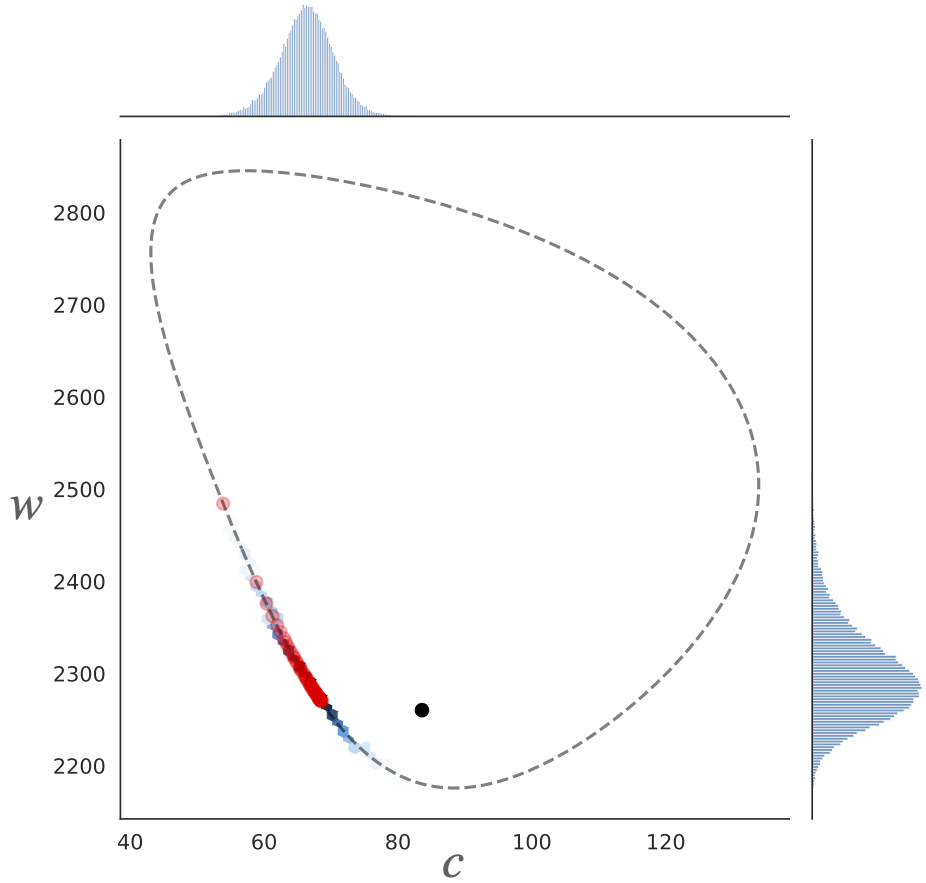}\label{FIG:KDEsc}}
  \caption{Jointplots and kernel density estimates of the tipped realizations of \eqref{EQ: CC EM} with $\epsilon=5$. The stable fixed point $z^*$ is denoted by the black circle and the unstable periodic orbit $\Gamma_u$ is denoted by the gray dashed curve. (a) Jointplot of points $(c_i,w_i)$ that escaped through $\Gamma_u$. We can see a clear region of exit points. (b) Jointplot of points $(c_i,w_i)$ from 65008 realizations that escaped through $\Gamma_u$ overlaid with trajectories of the subriver (red). (c) Jointplot of points $(c_i,w_i)$ from realizations that escaped through $\Gamma_u$ overlaid with where subriver exits $\Gamma_u$ (red points).}

\end{figure*}

This process results in converged distributions for exit location in both $c$ and $w$ along $\Gamma_u$ for this noise regime. Collecting the points $(c_i,w_i)$ from the paths that escaped, we see they fall on a specific part of $\Gamma_u$. In Figure \ref{FIG:KDEsa} see a jointplot of the exit locations respectively and notice that there is a distinct spot on $\Gamma_u$ where trajectories mostly exit. Additionally, see the overlay of the subriver and just the mouth of the subriver with the jointplot in Figure \ref{FIG:KDEsb} and Figure \ref{FIG:KDEsc} respectively. The majority of the subriver exits at the densest spot of the exit points.

Notice that besides the noisy realizations, these calculations do not depend on the noise strength. While vanishingly small noise predicts cycling, Freidlin-Wentzell theory can be used to show how cycling is actually resisted when noise strengths are larger. Of interest are the points on $\Gamma_u$ where the River trajectories cross as they escape, in which a subset of these trajectories, the subriver, lined up almost exactly with the Monte Carlo simulations. However, we currently still do not know what guides the noisy realizations, leading us to the next subsection.

\subsection{The Onsager-Machlup functional} \label{sec:OM}

To understand why the noisy trajectories escape via a specific area when escaping $\Gamma_u$, we need to calculate the energy needed to escape to higher order. This involves the Onsager-Machlup functional \cite{OM_Durr_Bach} which becomes relevant when the noise strength is away from the small noise limit. The Onsager-Machlup functional is used as a perturbation of the Freidlin-Wentzell functional and applied to trajectories in $\mathcal{R}$. 

\begin{figure}[ht]
    \centering
    \includegraphics[scale=.4]{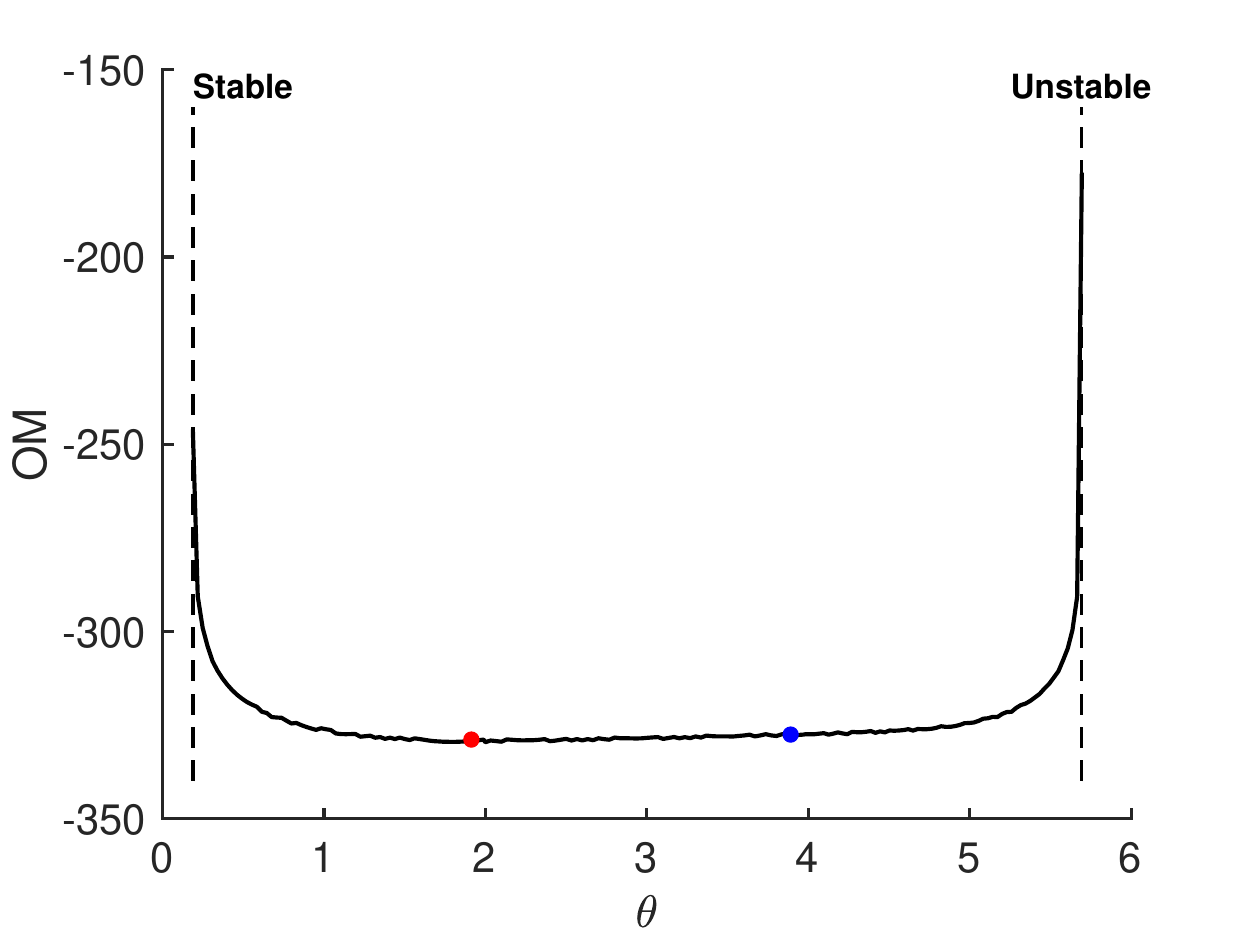}
    \caption{The value of $\eqref{eq:OM}$ for each trajectory in $\mathcal{R}$. The angles of $\mathcal{H}_1$ and $\mathcal{H}_2$ are denoted by the black dashed lines. The absolute minimum action value is given by the red circle, while the blue circle represents the corresponding action value for the pivot point.}
    \label{fig:OM}
\end{figure}

The Onsager-Machlup functional for a path $\phi$ on an interval $[t_0, t_f]$ is given by
\begin{equation}\label{eq:OM}
    OM[\phi]= \int_{t_0}^{t_f} |\dot{\phi}-P(\phi)|^2  + \epsilon^2 (\nabla \cdot P(\phi)) dt,
\end{equation}
where $P$ is the determinisitic vector field. The Onsager-Machlup functional penalizes trajectories that cycle around the periodic orbit. Since the heteroclinic orbits will wind around a neighborhood of $\Gamma_u$ infinitely many times, the perturbation that was added to the Freidlin-Wentzell functional implies the heteroclinic orbits cease to be global minimizers when away from the small noise limit. To find the global minimizer of \eqref{EQ: CC mpep eq}, we have to calculate this Onsager-Machlup term for each of our trajectories within the subriver. This process provides a selection mechanism to pick out a specific most probable escape path from the group of local minimizers: the subriver trajectories. For each trajectory in ${\mathcal{R}}$ we calculate \eqref{eq:OM}, and the results are shown in Figure \ref{fig:OM}. We find the absolute minimum action value, which corresponds to an angle of a point on $K$, leading us to the desired path.

\begin{figure}[ht]
  \centering
{\includegraphics[width=0.4\textwidth]{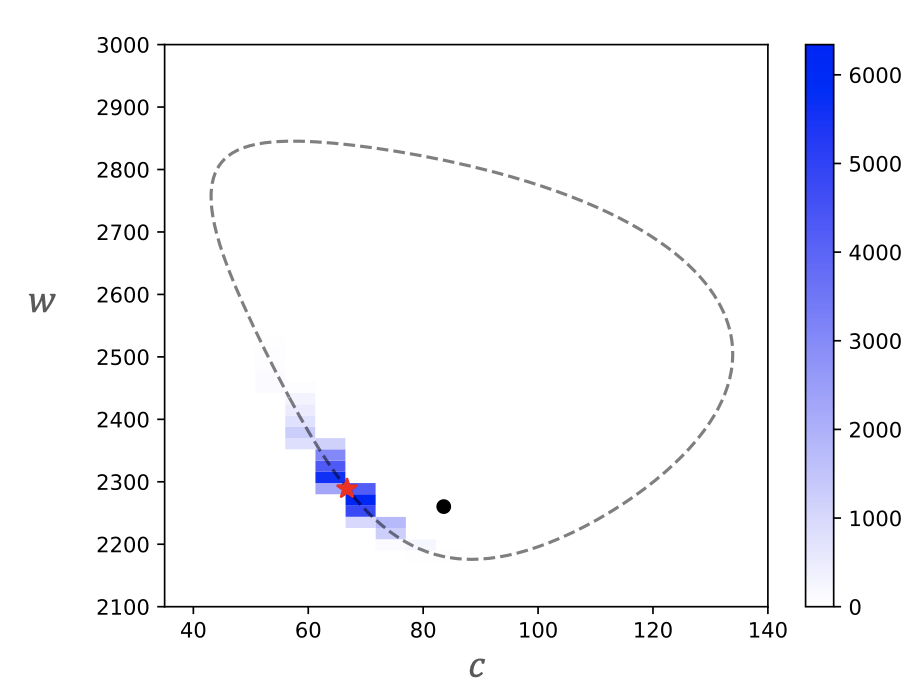}}
  \caption{Kernel density estimate of the tipped realizations of \eqref{EQ: CC EM} with $\epsilon=5$. The stable fixed point $z^*$ is denoted by the black circle and the unstable periodic orbit $\Gamma_u$ is denoted by the gray dashed curve. The KDE estimate of exit location (blue), overlaid with the exit location of the selected escape path (red star).}
\label{FIG:KDEs}
\end{figure}

Once we have the selected path, it needs to be verified with Monte Carlo simulations. A kernel density estimation will be used to determine how noisy trajectories exit $\Gamma_u$.
The chosen trajectory from the Onsager-Machlup functional will match the distribution of noisy trajectories all along the path, where this trajectory is acting as a guide for the realizations exiting $\Gamma_u$, as seen in Figure \ref{FIG:KDEs}.

\subsection{The role of the pivot point}
The hetroclinic orbits themselves never cross the periodic orbit. The fact that they cycle around inside the periodic orbit is at the heart of the cycling phenomenon discovered by Day \cite{day_exit_1996}, and elaborated upon by Berglund and Gentz \cite{berglund2004noise,berglund2014noise}. They do, however, delineate the orbits that do cross. Only the orbits with Maslov Index $0$ are expected to play a role in the most likely trajectories for exiting. These are the orbits 
which we have called the sub-river. Each of these orbits can be viewed as minimizers of the Freidlin-Wentzell functional with the constraint that the exit happen at a prescribed point on the periodic orbit $\Gamma$. Any given such minimizer will be at a local minimum because the cycling will generate infinitely many with, ironically, the functional value decreasing with extra cycling.  

Nevertheless, despite the preference for cycling as noise vanishes, a main point of this paper has been that moderate noise will favor less cycling. A natural question arises as to whether there is a lower limit on the cycling as noise increases. Our claim is that the pivot point plays a key role in determining this lower limit. This is due to its being the end-point of the exit points of trajectories with Maslov Index $0$. 

\begin{figure*}[ht]
  \centering
  \subfloat[]{\includegraphics[width=0.46\textwidth]{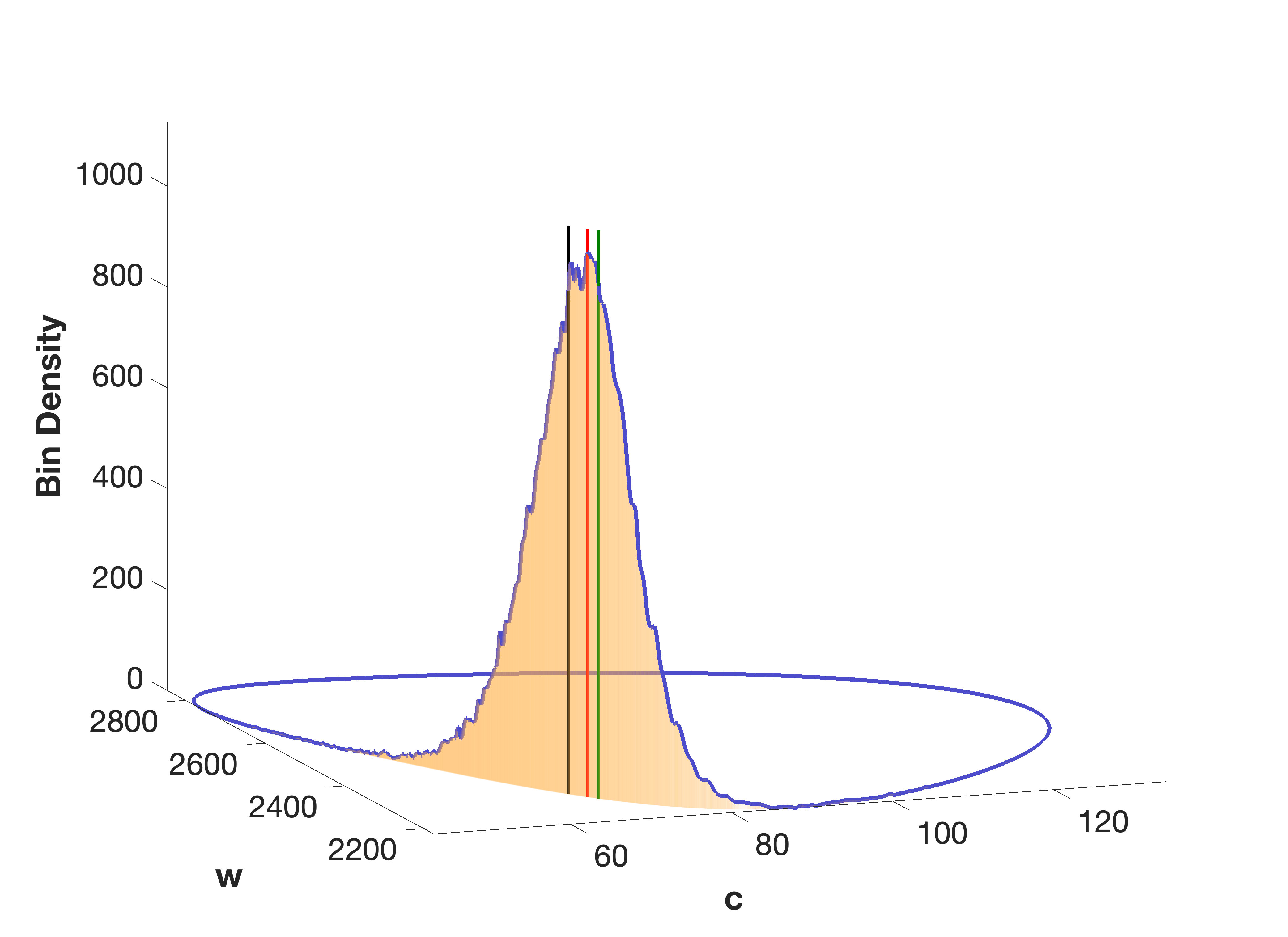}} 
  \hspace{-2mm}  
  \subfloat[]{\includegraphics[width=0.46\textwidth]{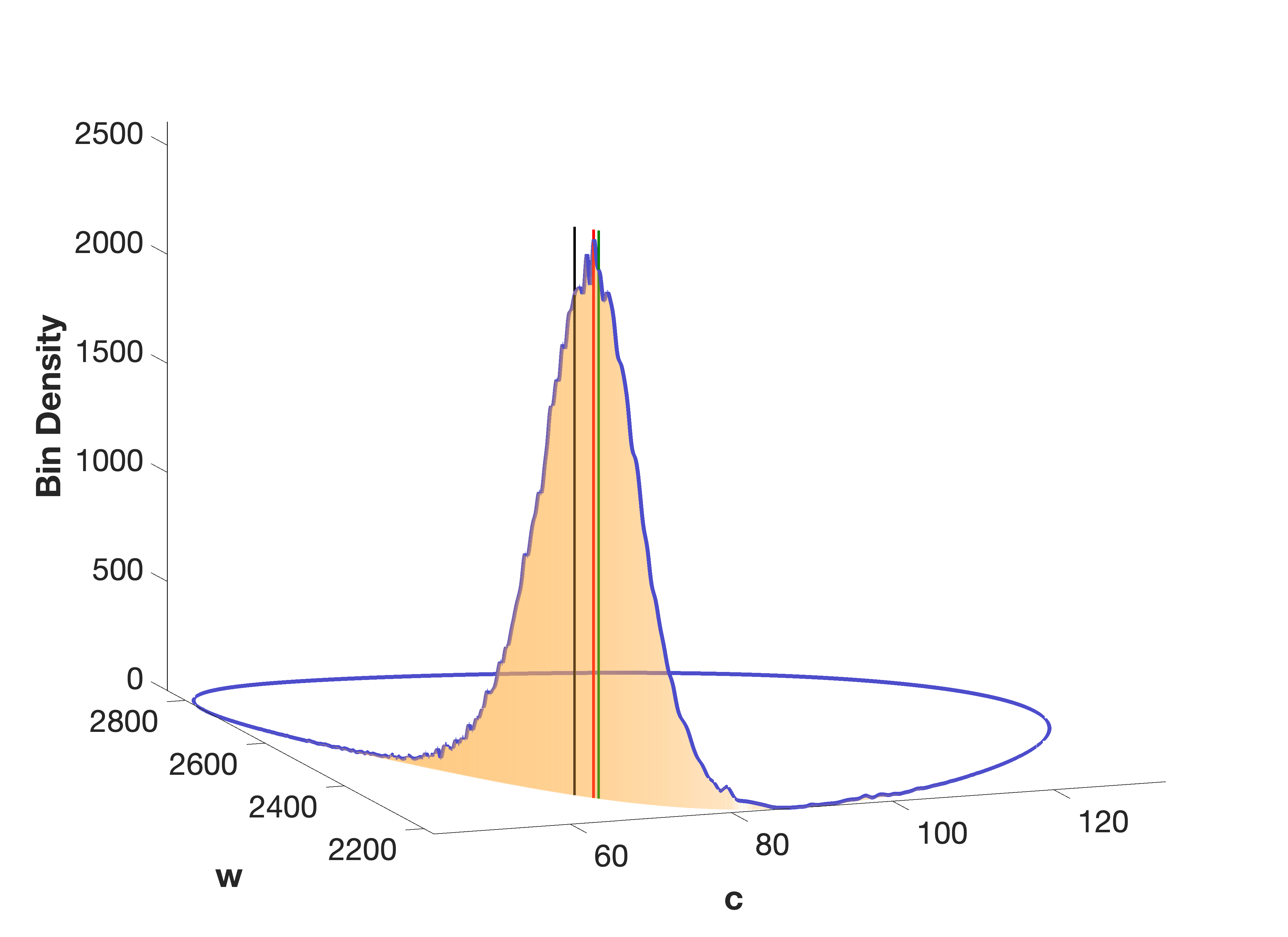}\label{fig:piv1}} 
  \vspace{-4mm} \\  
  \subfloat[]{\includegraphics[width=0.46\textwidth]{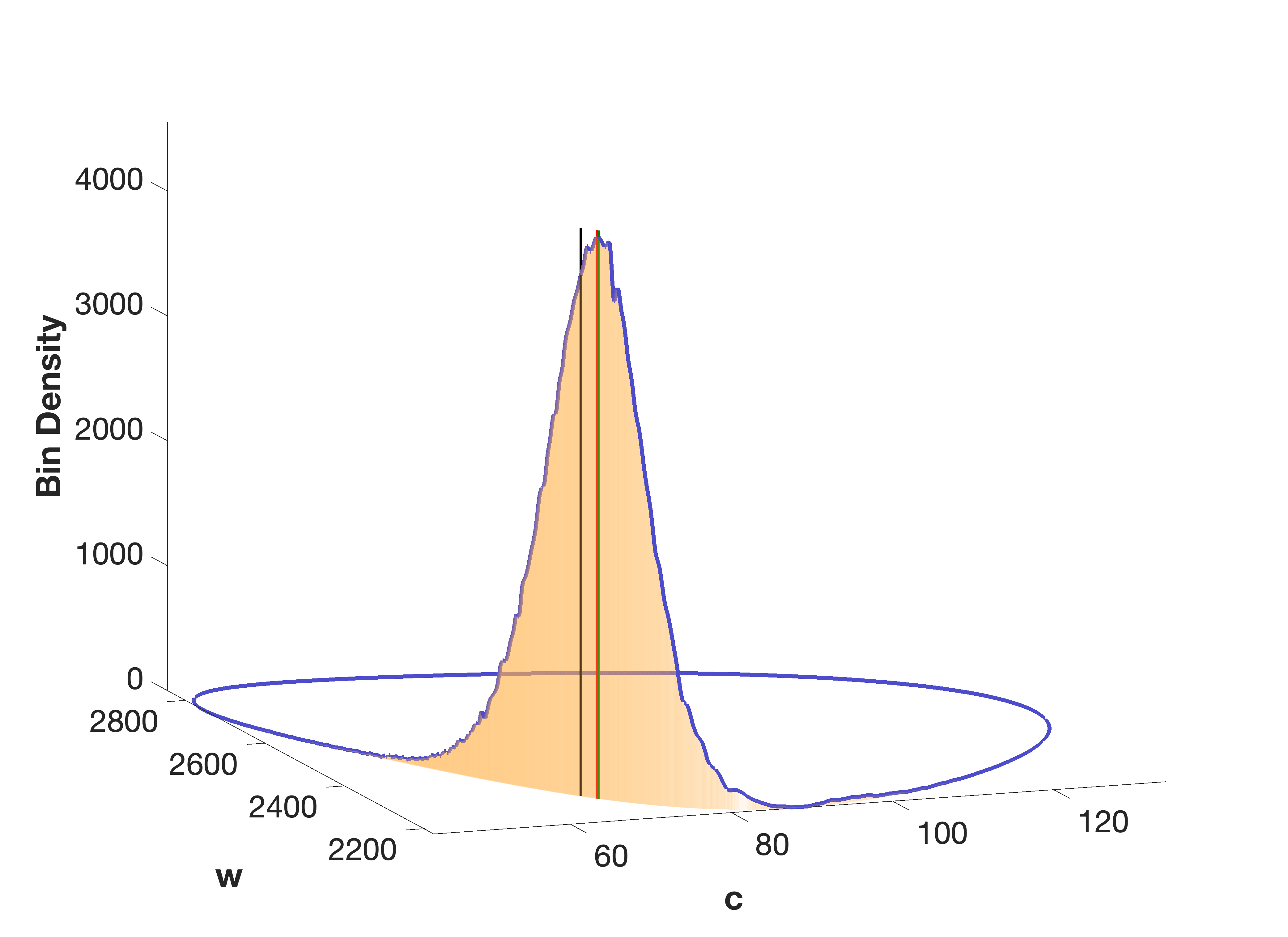}} 
  \hspace{-2mm}  
  \subfloat[]{\includegraphics[width=0.46\textwidth]{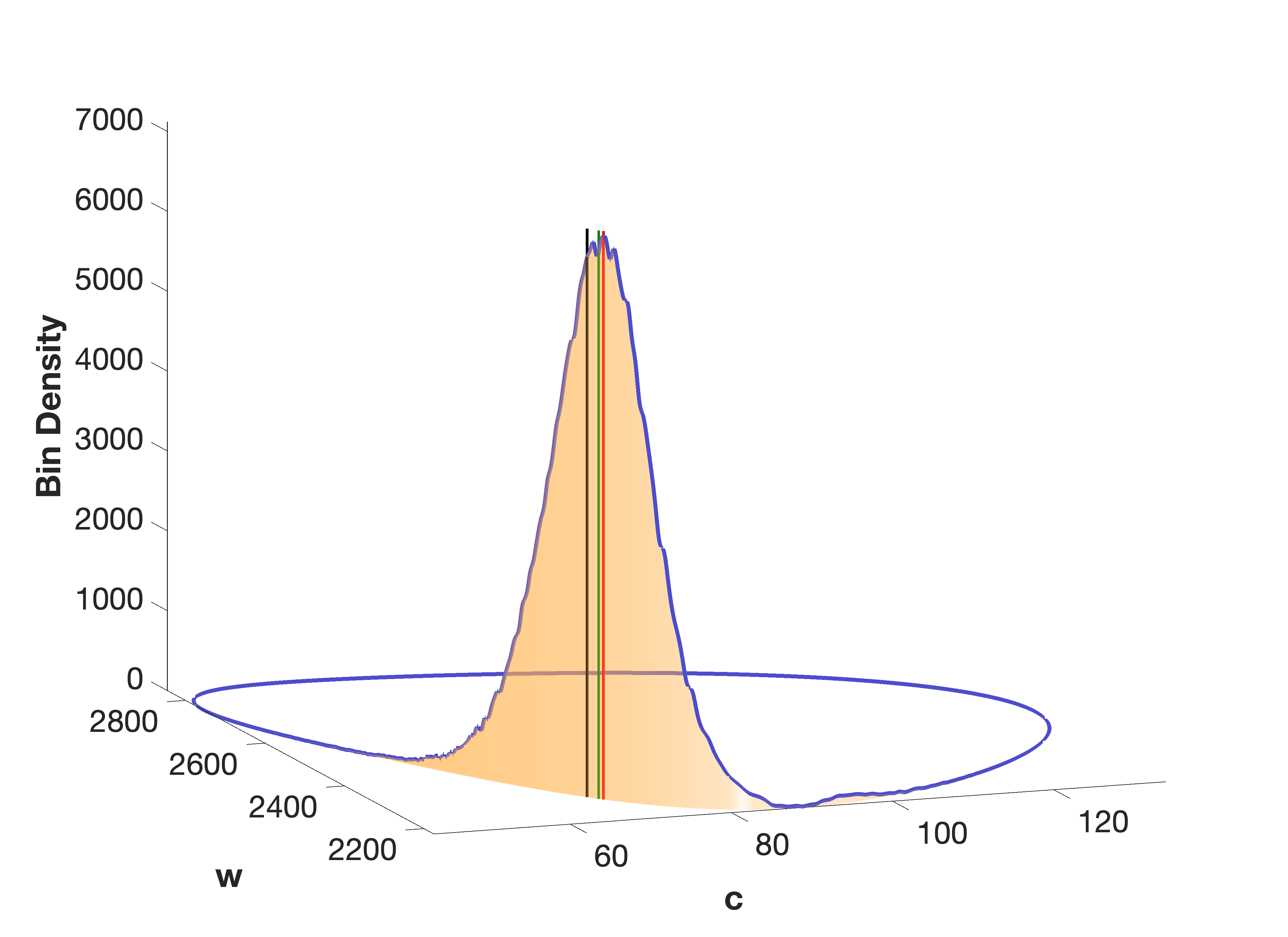}\label{fig:piv2}} 
  \caption{Probability exit distribution on the periodic orbit $\Gamma_u$ for various scenarios, where black represents the mean, red represents the mode, and green represents the pivot point: (a) $\epsilon=5$, 4\% tipping rate; (b) $\epsilon=5.5$, 9.3\% tipping rate; (c) $\epsilon=6$, 17\% tipping rate; (d) $\epsilon=6.5$, 26.7\% tipping rate.}
  \label{fig:Pivot-Action}
\end{figure*}

We ran a set of experiments with increasing noise and plotted the exit distributions from Monte-Carlo simulations. These are shown in Figure \ref{fig:Pivot-Action}. As the noise strength increases through the noted values, the mean and mode of the exit distribution get closer to the pivot point but appear to be blocked by it from rotating back further. 

The mode first passes the pivot point at the largest noise we took, namely $6.5$ which results in over $25\%$ of trajectories exiting. It remains, however, very close to the pivot point and the mean does not jump over the pivot point even at that relatively large level of noise, indicating that the bulk of the distribution remains short of the pivot point. 

\section{Discussion} \label{sec:discussion}

In this work, we conducted a detailed mathematical investigation of the susceptibility of a marine carbonate cycle model to rate-induced and noise-induced tipping from a stable fixed point, $z^* = (c^*,w^*) \in \mathbb{R}^2$, to a stable periodic orbit, $\Gamma_s$, in a bistable regime. The model, originally proposed by \citet{Rothman.2019}, captures the essential features of the marine carbonate system and exhibits rich dynamical behavior, including multiple stable states and hysteresis.
We first investigated the phenomenon of rate-induced tipping by making the CO$_2$ injection rate parameter a time-dependent ramping function. Through a rigorous bifurcation analysis and the computation of basins of attraction, we demonstrated that the system exhibits forward threshold instability. Specifically, we showed that if the ramping rate is sufficiently fast, the system trajectory can leave the basin of attraction of the stable fixed point before the fixed point itself loses stability via a bifurcation. This highlights the risk of critical transitions in the carbon cycle due to rapid anthropogenic forcing, even if the forcing remains below the static bifurcation threshold.

To study noise-induced tipping, we added stochastic white noise perturbations to the deterministic system and sought to identify the most probable path of escape from the stable fixed point to the unstable periodic orbit that forms the basin boundary. Our computational results revealed that for small but non-vanishing noise levels, system realizations concentrate their escape through a localized region on the unstable periodic orbit. This differs from the infinite cycling around the orbit predicted by large deviation theory in the small noise limit.

To uncover the geometric structure underlying the observed escape behavior, we computed the unstable manifold of the fixed point $W^u(z^*)$ and the stable manifold of the unstable periodic orbit $W^s(\Gamma_u)$. We defined sets of trajectories called the River, $\mathcal{R}$, and subriver, $\tilde{\mathcal{R}}$, by looking at intersections of these manifolds following the work from \citet{Emman_IVDP}. The subriver consists of trajectories that are local minimizers of the Freidlin-Wentzell action functional. Through kernel density estimation of Monte Carlo simulations of the stochastic system, we confirmed that the escape location of realizations aligns with where the subriver trajectories cross the unstable periodic orbit $\Gamma_u$.
To identify the global minimizer of the Freidlin-Wentzell action functional from among the local minimizers in the subriver under small but finite noise, we employed the Onsager-Machlup functional as a perturbation. This functional penalizes trajectories that wind around $\Gamma_u$ multiple times by adding a term that grows with the amount of cycling. By minimizing the Onsager-Machlup functional over the subriver trajectories, we were able to select a specific most probable escape path. We validated this path by showing it very closely approximates the mode of the escape location distributions obtained from Monte Carlo simulations, see Figure \ref{FIG:KDEs}.

Our results demonstrate that the marine carbonate cycle model is susceptible to both rate-induced and noise-induced tipping in realistic parameter regimes, even when using standard mathematical formulations of these phenomena. The rate-induced tipping analysis underscores the danger of the system being pushed out of a stable operating state by rapid anthropogenic increases in CO$_2$ emissions, before a bifurcation threshold is reached. The noise-induced tipping analysis reveals the most likely routes for stochastic escape from a metastable operating state and shows how cycling of escape trajectories around the basin boundary is resisted under small but realistic noise levels.

This work advances the methodology for rigorously investigating tipping susceptibility in climate-relevant models using techniques from multiple areas of applied mathematics. We leveraged tools from geometric singular perturbation theory to identify forward threshold instability, Hamiltonian dynamical systems theory to compute invariant manifolds and connecting orbits, large deviation theory to define most probable escape paths, and Monte Carlo simulations to sample stochastic escape trajectories.

The mathematical approach developed here could potentially be extended to study tipping in higher-dimensional models of the climate system. Some promising avenues for future work include:

\begin{enumerate}
\item Analyzing more complex and realistic models of the carbon cycle that include additional physical, chemical, and biological processes, as well as coupling to other climate system components like the atmosphere, ocean, and land surface.
\item Investigating rate-induced tipping under more gradual parameter drift scenarios, such as those corresponding to different anthropogenic CO$_2$ emissions pathways.
\item Studying noise-induced tipping under non-white, multiplicative, and non-Gaussian noise assumptions that better reflect the stochastic variability present in the real climate system.
\item Exploring the interplay between rate-induced and noise-induced tipping by making time-varying parameters also fluctuate stochastically.
\item Extending the invariant manifold and most probable path computational methods to systems with higher dimensionality, e.g. by leveraging numerical continuation techniques.
\item Calibrating model parameters and noise levels using observational data and paleoclimate proxy records to obtain more quantitative estimates of tipping risks.
\item Developing early warning indicators for rate and noise-induced tipping based on the dynamical features identified in this analysis, such as forward threshold instability combined with subriver escape trajectories in non-gradient systems.
\end{enumerate}

In conclusion, this paper demonstrates the power of a mathematically exhaustive approach for analyzing the susceptibility of a climate-relevant model to multiple tipping phenomena. Our results show the marine carbonate system is at risk of undergoing critical transitions due to both rapid anthropogenic forcing and stochastic variability, even in a relatively simple model. By extending this approach to more comprehensive models, we can work towards identifying and quantifying tipping risks in the global climate system. A deeper mathematical understanding of these risks is critical for informing effective strategies to mitigate and adapt to climate change in the face of both gradual and abrupt shifts. Future work should focus on further developing the mathematical theory and computational tools needed to assess tipping in complex, high-dimensional stochastic systems, as well as applying these methods to actionable climate change decision-making.

\section*{Acknowledgement} 

Katherine Slyman was supported
by the NSF grant RTG: DMS-2038039. Emmanuel Fleurantin was supported by the NSF grant DMS-2137947. Christopher K.R.T. Jones was supported by the Office of Naval Research under grant N00014-24-1-2198.

\section*{Author Declarations}

The authors have no conflicts to disclose.

\nocite{*}
\bibliographystyle{apsrev4-1}
\bibliography{BIBBIB}

\end{document}